%% file: main.tex
\documentclass[conference]{IEEEtran} 
\usepackage[numbers,sort&compress]{natbib} 

\input{math_commands.tex}

\usepackage{hyperref}
\usepackage[nameinlink,noabbrev]{cleveref}
\usepackage{url}

\usepackage[utf8]{inputenc} 
\usepackage[T1]{fontenc}    
\usepackage{url}            
\usepackage{booktabs}       
\usepackage{amsfonts}       
\usepackage{graphicx}
\usepackage{pdfpages} 
\usepackage{multirow}
\usepackage{nicefrac}       
\usepackage{microtype}      
\usepackage{xcolor}         
\usepackage{enumitem}
\usepackage{tcolorbox}
\tcbuselibrary{most}
\tcbuselibrary{skins}
\usepackage{fontawesome5}
\usepackage{subcaption}


\definecolor{promptblue}{RGB}{41, 128, 185}
\definecolor{promptgray}{RGB}{248, 249, 250}
\definecolor{jsonbg}{RGB}{250, 250, 250}

\definecolor{promptgray}{RGB}{248,249,250}
\definecolor{promptblack}{RGB}{33,37,41}
\definecolor{promptblue}{RGB}{0,123,255}
\definecolor{nowatermarkgray}{RGB}{248,249,250}
\definecolor{nowatermarkborder}{RGB}{108,117,125}
\definecolor{attackredbg}{RGB}{255,245,245}
\definecolor{attackredborder}{RGB}{170,35,35}
\definecolor{jsonbg}{RGB}{240,240,240}
\definecolor{jsonbg}{RGB}{240,240,240}

\newcommand{\panelbox}[3]{%
\begin{tcolorbox}[
    enhanced,
    colback=#1,
    colframe=#2,
    boxrule=1.5pt,
    arc=8pt,
    title={#3},
    fonttitle=\bfseries\large,
    coltitle=white,
    attach boxed title to top center={yshift=-2mm},
    boxed title style={colback=#2, arc=3pt},
    drop shadow={black!20!white},
    left=12pt,
    right=12pt,
    top=18pt,
    bottom=12pt,
    height=7.8cm,
    valign=top
]
}

\def \ie {\emph{i.e.}}
\def \eg {\emph{e.g.}}

\usepackage{algorithm}
\usepackage{algorithmic}
\usepackage{amsmath,amssymb}

\usepackage{amsthm}          
\newtheorem{assumption}{Assumption}[section] 

\crefname{assumption}{assumption}{assumptions}
\Crefname{assumption}{Assumption}{Assumptions}

\newtheorem{theorem}{Theorem}

\newif\ifsatmlfinal
\satmlfinaltrue 

\title{Mitigating Watermark Forgery in Generative Models via Randomized Key Selection}

\ifsatmlfinal
\author{%
  Toluwani Aremu\textsuperscript{1} \quad
  Noor Hussein\textsuperscript{3} \quad
  Munachiso Nwadike\textsuperscript{1} \quad
  Samuele Poppi\textsuperscript{1} \\
  Jie Zhang\textsuperscript{2} \quad
  Karthik Nandakumar\textsuperscript{1,3} \quad
  Neil Gong\textsuperscript{4} \quad
  Nils Lukas\textsuperscript{1} \\
  \small
  \textsuperscript{1}\textit{MBZUAI, UAE} \quad
  \textsuperscript{2}\textit{A*STAR, Singapore} \quad
  \textsuperscript{3}\textit{Michigan State University, USA} \quad
  \textsuperscript{4}\textit{Duke University, USA}
}
\else
\author{\IEEEauthorblockN{Anonymous Authors}}
\fi

\begin{document}

\maketitle

\begin{abstract}
Watermarking enables GenAI providers to verify whether content was generated by their models.
A watermark is a hidden signal in the content, whose presence can be detected using a secret watermark key.
A core security threat are forgery attacks, where adversaries insert the provider's watermark into content \emph{not} produced by the provider, potentially damaging their reputation and undermining trust.
Existing defenses resist forgery by embedding many watermarks with multiple keys into the same content, which can degrade model utility.
However, forgery remains a threat when attackers can collect sufficiently many watermarked samples.
We propose a defense with a sample-count-independent upper bound on forgery success for blind attackers, conditional on key-symmetric, independent detector outcomes.
Our scheme does not further degrade model utility.
We randomize the watermark key selection for each query and accept content as genuine only if a watermark is detected by \emph{exactly} one key.
Unlike cryptographic watermarks that rely on computational hardness assumptions and require designing new watermarking schemes from scratch, our method can be applied to any existing watermarking method to improve its forgery resistance.
We focus on text watermarking, but our defense is modality-agnostic, since it treats the underlying watermarking method as a black-box. To show this, we include a preliminary study on image watermarking using Tree-Ring.
Separately from this conditional guarantee, we empirically observe that, at $r=4$ keys, harmful-text forgery success drops from as high as $87\%$ with a single key to as low as $1\%$ against the adaptive blind attackers that we evaluate, at negligible computational overhead; a preliminary image study shows a reduction from $100\%$ to $2\%$.
\end{abstract}

\input{parts/1_intro}
\input{parts/2_background}
\input{parts/3_method}
\input{parts/4_exp}
\input{parts/5_summary}

\bibliography{iclr2027_conference}
\bibliographystyle{IEEEtranN} 

\appendices
\input{appendix/1_rel}

\input{appendix/2_utility}
\input{appendix/3_ablation}

\input{appendix/4_attacks}
\clearpage

\end{document}

%% file: math_commands.tex
\usepackage{amsmath,amsfonts,bm}

\def\eqref#1{equation~\ref{#1}}

\def\1{\bm{1}}

\DeclareMathAlphabet{\mathsfit}{\encodingdefault}{\sfdefault}{m}{sl}
\SetMathAlphabet{\mathsfit}{bold}{\encodingdefault}{\sfdefault}{bx}{n}



%% file: parts/1_intro.tex
\section{Introduction} 
\label{sec:introduction}

Large generative models are often trained by a few providers and consumed by millions of users. 
They produce high-quality content~\citep{bubeck2023sparks, grattafiori2024llama}, which can undermine the authenticity of digital media~\citep{he2024mgtbench} when misused, such as spreading online spam or disinformation.
Watermarking enables the attribution of generated content by hiding a message that is detectable using a secret watermarking key~\citep{kirchenbauer2023watermark, zhao2024provable, christ2024undetectable}.

A security property allowing watermarks to function as digital signatures is called \emph{forgery-resistance}, which means that embedding a watermark can only be done with knowledge of the secret key. 
A threat to providers are forgery attacks~\citep{gu2024on, jovanovic2024watermarkstealing}, where adversaries try to insert an inauthentic watermark into content not generated by the provider to falsely attribute it to the provider's LLM. 
For example, an attacker could use an open model to generate harmful content denying historical events or promoting violence, then add the provider's watermark to falsely implicate the provider's model as the source.
Such attacks are particularly damaging because they exploit the provider's reputation while potentially exposing them to legal liability and regulatory scrutiny \citep{eu_2024, CA_SB942_2024}. 

Existing defenses against forgery have significant limitations. 
Statistical detection methods~\citep{gloaguen2024discovering} attempt to distinguish genuine from forged watermarks by analyzing patterns such as token distributions, n-gram frequencies, or textual artifacts. 
Another approach is to rotate keys after revealing $N$ samples, but this introduces key management complexity and leaves the optimal choice of $N$ unclear. 
The provider could also embed watermarks with multiple keys into the same content~\citep{Pang2024NoFL}.
However, \citet{jovanovic2024watermarkstealing, zhao2024sok} show that resisting forgery attacks when the adversary collects $N$ or more watermarked samples remains an open problem.

\begin{figure}[t!]
    \centering
    \includegraphics[width=\linewidth]{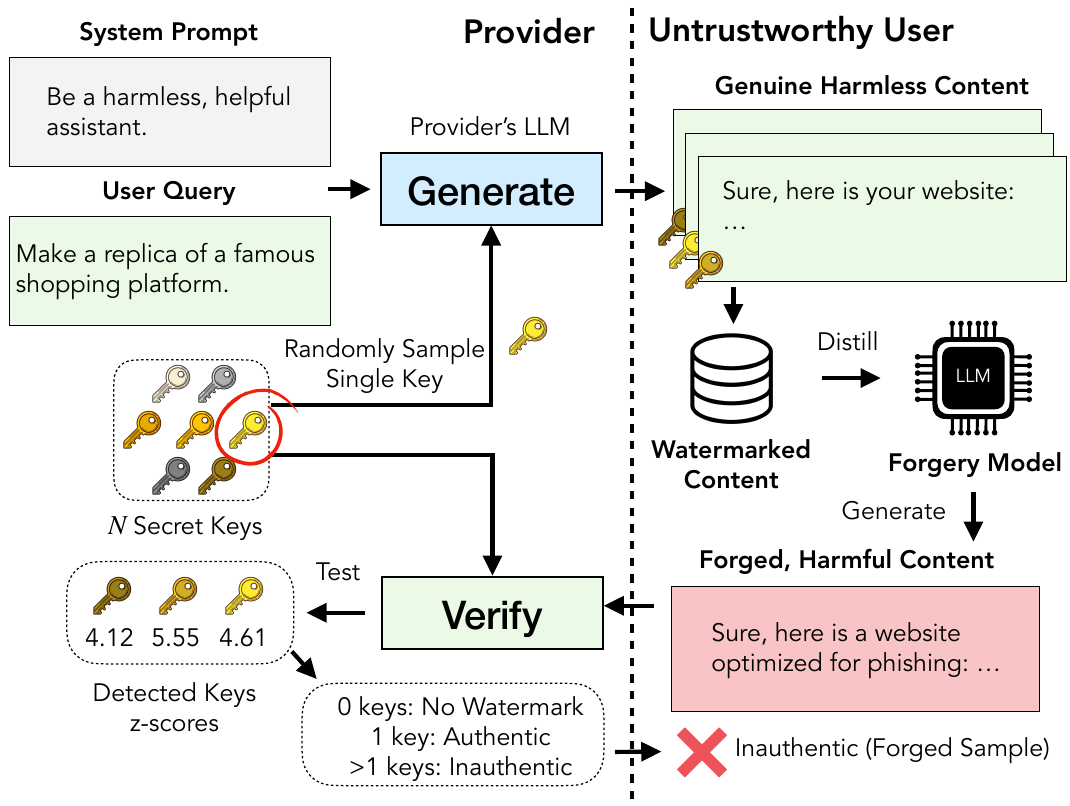}
    \caption{An overview of forgery attacks and our proposed randomization strategy for watermarking key selection to improve forgery-resistance.}
    \label{fig:eyecatcher}
\end{figure}

\noindent \textbf{Our Method.} \Cref{fig:eyecatcher} illustrates the core idea of our method. 
During generation, our method randomizes key selection from a pool of keys $\mathcal{K}$ for each user query, \emph{using exactly one key per sample}, identical to standard single-key watermarking.
To detect the presence of a watermark, it detects whether a watermark can be detected with any of the keys in $\mathcal{K}$ and accepts content as genuinely watermarked only if the presence of \emph{exactly} one key is statistically significant. 
Otherwise, if no key is detected (0 keys), the content was not generated by the provider's model and if multiple keys are detected ($\geq 2$ keys), the content has been forged. In particular, samples containing multiple detected keys are \emph{always rejected as non-genuine}.
\textbf{Note }that our goal is to provide a practical security layer against forgery for existing watermarking systems, rather than redesigning them.

\noindent \textbf{Advantages of our Method.} Our defense admits a conditional upper bound on forgery success that is independent of the number $N$ of watermarked samples revealed to the attacker.
The bound applies to \emph{blind attackers} who cannot reliably separate samples by key, provided their forged outputs induce independent detector outcomes with a common detection probability across keys.
Our method inherits the underlying watermark's detectability and robustness properties, without further degrading the model's utility, improving over prior multi-key approaches evaluated by \citet{Pang2024NoFL} and \citet{kirchenbauer2023watermark}.
Moreover, a failed forgery is not merely rejected: triggering two or more keys is positive evidence that the content was manipulated, which single-key detection cannot provide (\Cref{sec:discussion}).
We focus on text and image, but our method is \textbf{modality-agnostic} as it treats the underlying watermarking method as a black box.
Under these assumptions, maximizing over the attacker-induced per-key detection probability $p$ gives an upper bound of $(1 - 1/r)^{r-1}$, attained at $p=1/r$. This conditional bound is \textbf{independent of the number of samples} available to the attacker; it does not establish negligible cryptographic forgery probability.
Empirically, at $r=4$ keys, our method reduces harmful-text forgery success from as high as $87\%$ with a single key to as low as $1\%$ against adaptive blind attackers; a preliminary image study shows a reduction from $100\%$ to $2\%$.
%

\noindent \textbf{Contributions.} We: \textbf{(i)} propose a simple, black-box randomized key selection strategy that significantly improves resistance to watermark forgery across all surveyed attacks.
\textbf{(ii)} provide the first theoretical bound on forgery success that is \emph{independent of the number of samples} available to blind attackers, conditional on key-symmetric, independent detector outcomes.
\textbf{(iii)} empirically validate our approach on four text watermarking schemes, reducing harmful-text forgery success of adaptive blind attackers to as low as $1\%$ at $r=4$, with a preliminary image study on Tree-Ring.
\textbf{(iv)} release our implementation as open-source code.

%% file: parts/2_background.tex
\section{Background} \label{sec:background}
\noindent \textbf{Language Modeling.}
LLMs generate text by predicting tokens based on previous context. Formally, for vocabulary $\mathcal{V}$ and sequence $x = (x_1, x_2, \ldots, x_n)$ where $x_i \in \mathcal{V}$, an LLM defines:
\begin{equation}
p(x) = \prod_{i=1}^{n} p(x_i | x_{<i}),
\end{equation}
where $x_{<i} = (x_1, \ldots, x_{i-1})$ are the tokens in the model's context.

\noindent \textbf{LLM Content Watermarking.} 
Content watermarking provides a mechanism for attributing generated content to specific models, enabling accountability and misuse detection.
Watermarks are hidden signals in generated content that can be detected using a secret watermarking key.
Watermarking methods are formalized by two algorithms:
(1) an embedding algorithm $\operatorname{Watermark}^\mathcal{M}_{k}(\pi) \rightarrow x$ that produces watermarked content using a private watermarking key $k$, model $\mathcal{M}$, prompt $\pi$, and
(2) a detection algorithm $\operatorname{Detect}_{k}(x)$ that outputs the statistical significance of the presence of a watermark~\citep{zhao2024sok}.
A parameter $\tau\in \mathbb{R}$ represents the minimum decision threshold. 

A watermarking method is \emph{unforgeable} if it is computationally infeasible for an adversary who does not know the secret key to produce content that passes the watermark detection test ~\citep{christ2024undetectable, christ2024pseudorandom, gunn2025undetectable}.
Formally, unforgeability requires that for all polynomial-time algorithms $\mathcal{A}$, the probability that $\mathcal{A}$ can generate content $x$ such that $\operatorname{Detect}_{k}(x) > \tau$ while $x$ was not produced by the watermarked model is negligible. That is, a watermark is \emph{unforgeable} if for all security parameter $\lambda$ and polynomial-time algorithms $\mathcal{A}$ if,
\vspace{-.05cm}
\begin{align}
    \Pr_{\substack{k \\ x \gets \mathcal{A}^{\operatorname{Watermark}^{\mathcal{M}}_{k}(1^{\lambda}, k)}}}
    \hspace{-3.5em}
    \left[
        \operatorname{Detect}_{k}(x) > \tau \text{ and } x \notin \mathcal{Q}
    \right] 
    \leq \mathsf{negl}(\lambda),
\end{align}
where $\mathcal{Q}$ denotes the set of responses obtained by $\mathcal{A}$ on its queries to the watermarked model. 
This property ensures that watermarks can be reliably used for content attribution, preventing malicious actors from falsely attributing harmful content to a specific model provider.
We refer to \Cref{sec:related_work} for more details on the surveyed watermarking methods and forgery attacks.

\section{Threat Model} \label{sec:threat}

Consider a GenAI model provider that uses watermarking to attribute content generated by their models. 
The provider is responsible and deploys mechanisms to prevent their models from generating \emph{harmful} content, such as misinformation, hateful content, or malware~\citep{bai2022constitutional}.  
A threat to the provider is untrustworthy users who generate harmful content without using the provider's service and inject the provider's watermark, allowing them to impersonate and falsely accuse the provider, eroding trust in the attribution system.  
The provider needs methods to mitigate forgery.

\noindent \textbf{Provider's Capabilities.} 
(\textit{Secrecy}) We assume the provider can store multiple watermarking keys securely and 
(\textit{Model Deployment}) has full control over their model's deployment, \eg, they can process generated content prior to its release. 
Additionally, (\textit{Safety Filters}) we assume the provider implements effective safety mechanisms that prevent a model from generating content from a set $\mathcal{H}\subseteq \mathcal{V}^*$ considered to be harmful. 
While many existing safety mechanisms are still vulnerable to jailbreak attacks~\citep{wei2023jailbroken, poppi2025towards}, more advanced defenses are being developed, and in the near future, it may be infeasible (\eg, due to high cost) to jailbreak frontier LLMs.

\noindent \textbf{Attacker's Capabilities.} 
(\textit{Model Access}) The attacker has black-box API access to the provider's model which they can use to collect watermarked samples using any prompt. 
(\textit{Adaptive}) We assume an adaptive attacker who knows which watermarking method is used by our defender (incl. hyper-parameters), but they do not know the secret watermarking keys \citep{lukas2024leveraging, diaa2024optimizing}.
We further distinguish between (i) \emph{blind} attackers who cannot easily separate watermarks detected by different keys, and (ii) \emph{informed} attackers who are given $N$ labeled watermarked samples. 
Blindness is a capability limit: the absence of key labels alone does not imply that an attacker cannot separate samples by key, nor does it imply independence of detector outcomes.
(\textit{Private Detector}) 
The watermark detection API is not accessible to the attacker.  
(\textit{Query Budget}) Unlike prior works \citep{jovanovic2024watermarkstealing}, our attacker is unrestricted in the number of queries they can make. 

\noindent \textbf{Attacker's Goal.} The attacker's goal is to generate harmful content $x$ that (i) the provider would refuse to generate ($x\in \mathcal{H}$) and (ii) is accepted as genuine by our multi-key detector, meaning that exactly one key is detected ($s(x)=1$).
We do not explicitly define `harmful' content, but instead use a pre-existing instruction-tuned model and consider content harmful if the model refuses to generate it with a high probability. We do not consider robustness or partial-attribution attacks in this work, as they correspond to a different detection objective.

\subsection{Security Game}  
We formalize watermark forgery as a game between a challenger (the provider) and an adversary $\mathcal{A}$.
\begingroup
Let $\mathcal{M}$ be the provider's language model and $\mathcal{K}=\{k_1,\ldots,k_r\}$ its secret watermarking key set. For each prompt $\pi$, the provider independently samples $k\sim\operatorname{Uniform}(\mathcal{K})$ and returns $x\sim\operatorname{Watermark}^{\mathcal{M}}_k(\pi)$ without revealing the selected key. The adversary can adaptively make $N$ queries to obtain pairs $\{(\pi_i,x_i)\}_{i=1}^N$. Given a harmful target prompt $\pi^\star$, the adversary outputs a forged sample $\hat{x}^\star$. Define the number of detected keys as
\[
 s(x)=\sum_{j=1}^{r}\mathbf{1}\{\operatorname{Detect}_{k_j}(x)>\tau\}.
\]
The adversary wins if the forged sample is accepted as genuine ($s(\hat{x}^\star)=1$) and is harmful ($\hat{x}^\star\in\mathcal{H}$). Its forgery success probability is:
\begin{align}
\text{Adv}^{\text{forge}}_{\mathcal{A}} = \Pr[s(\hat{x}^\star)=1 \land \hat{x}^\star \in \mathcal{H}].
\end{align}
Cryptographic unforgeability would require negligible success for all efficient adversaries. We do not establish that property for our wrapper: our contribution is a sample-count-independent conditional bound for blind attackers and empirical mitigation of evaluated forgery attacks.
\par\endgroup
We focus on heuristic watermarks (\eg, KGW, Unigram, Tree-Ring), which are widely deployed in practice due to their simplicity, low generation latency, and robustness, but lack the cryptographic unforgeability guarantees of~\citet{christ2024pseudorandom}. As stated in \Cref{sec:introduction}, our goal is to add a layer of statistical forgery resistance to these practical schemes.

%% file: parts/3_method.tex
\section{Conceptual Approach} 
\label{sec:conceptual_approach}

\Cref{alg:multi-key-watermarking} randomly samples a watermarking key from a set of keys $\mathcal{K}$ and uses it to generate watermarked content.
During detection, the provider runs per-key tests and applies a common threshold $\tau$ chosen via the Šidák correction to control the family-wise error rate $\alpha_{\mathrm{fw}}$ across $r=|\mathcal{K}|$ keys.
We declare \textbf{genuine} if \emph{exactly one} key is detected, \textbf{forgery} if two or more keys are detected, and \textbf{not ours} if no key is detected.
We refer to \Cref{sec:theoretical_analysis} for more information on the calibration.
\begin{algorithm}[ht]
\caption{Our Forgery Detection Algorithm}
\label{alg:multi-key-watermarking}
\begin{algorithmic}
\REQUIRE Prompt $\pi$, model $\mathcal{M}$, key set $\mathcal{K}=\{k_1,\dots,k_r\}$, family-wise error $\alpha_{\mathrm{fw}}$, null CDF $F_{0}$ of the per-key statistic
\ENSURE Watermarked response $x$ (Generation), Decision (Detection)
\vspace{1mm}
\STATE \textbf{Generation:}
\STATE $k \sim \mathcal{K}$ \COMMENT{Randomly sample a key}
\STATE $x \gets \operatorname{Watermark}^{\mathcal{M}}_{k}(\pi)$ \COMMENT{Watermarked response}
\RETURN $x$
\vspace{2mm}
\STATE \textbf{Detection:}
\STATE $\alpha \gets 1-(1-\alpha_{\mathrm{fw}})^{1/r}$ \COMMENT{Šidák per-key level, cf.\ \eqref{eq:sidak}}
\STATE $\tau \gets F_{0}^{-1}(1-\alpha)$ \COMMENT{Common threshold; fixed per \Cref{sec:threshold}}
\STATE $s \gets 0$
\FOR{$i = 1$ \TO $r$}
   \STATE $T_i \gets \operatorname{Detect}_{k_i}(x)$ 
   \STATE $Z_i \gets \mathbf{1}\{T_i>\tau\}$
   \STATE $s \gets s + Z_i$
\ENDFOR
\RETURN $s$ \COMMENT{$s=0$ (not ours), $s=1$ (ours), $s>1$ (forgery)}
\end{algorithmic}
\end{algorithm}

The idea of \Cref{alg:multi-key-watermarking} is that an attacker who collects watermarked samples \citep{jovanovic2024watermarkstealing, gu2024on} to use for their forgery process cannot simply distill without learning the statistical signals from all watermarks, since each sample contains a different watermark which is unknown to the attacker. 
An attacker that distills our watermark inadvertently 'poisons' their forgery attack with watermarks from different keys which the provider can detect. 
When attackers fail to identify whether the samples were generated with different keys, they will learn a mixture of the watermarks in their attack. 
In the next section, we theoretically analyze the forgery resistance of our method.

\subsection{Analyzing Forgery Resistance}
\label{sec:theoretical_analysis}

For any content $x$ and key $k_i\in \mathcal{K}$ define the indicator
\begin{align}
  Z_i(x)=\mathbf{1}\!\bigl\{\mathsf{Detect}_{k_i}(x)>\tau\bigr\}.
\end{align}
which represents outcome of the key-specific detector with threshold~$\tau$. 
The global null hypothesis $H_0$ is ``$x$ is not watermarked by any key.'' and for any $1\leq j \leq r$, $H_{1,j}$ is ``$x$ is watermarked with $k_j$''. 

\begin{assumption}
\label{ass:model}
There exist constants $\alpha\in(0,1)$ and $\beta\in(0,1]$ such that:
\begin{enumerate}[label=(\roman*),leftmargin=*,itemsep=0mm,topsep=0mm]
\item\label{ass:fpr} Under $H_0$, $\Pr[Z_i=1]=\alpha$ for all $i$.
\item\label{ass:indep0} Under $H_0$, $\{Z_i\}_{i=1}^r$ are mutually independent.
\item\label{ass:tpr} Under $H_{1,j}$, $\Pr[Z_j=1]=\beta$ and for $i\neq j$, $\Pr[Z_i=1]=\alpha$.
\item\label{ass:indep1} Under $H_{1,j}$, $Z_j$ is independent of $\{Z_i\}_{i\neq j}$ and $\{Z_i\}_{i\neq j}$ are mutually independent.
\end{enumerate}
\end{assumption}

\Cref{fig:wm_ana} provides empirical evidence of low cross-key interference and separation between target and non-target detector scores for the evaluated KGW-SelfHash and Unigram schemes. These marginal distributions do not establish joint independence or validate family-wise error control under the multi-key threshold. Independence in \Cref{ass:model} remains an assumption requiring separate joint-statistic validation.
   
\subsubsection{Calibrating the per-key threshold {$\tau$}}
\label{sec:threshold}

Let $T_i(x)$ be the raw statistic for key $k_i$ and $F_{0}$ its CDF under $H_0$.
Setting $\tau$ fixes the per-key FPR $\alpha(\tau)=1-F_{0}(\tau)$.
Under the null-independence assumption in \Cref{ass:model}, a non-watermarked $x$ triggers any key with probability $1-(1-\alpha)^r$.
To enforce a global budget $\alpha_{\mathrm{fw}}\in(0,1)$ we use \cite{vsidak1967rectangular} correction:
\begin{align}
\label{eq:sidak}
  \alpha \;:=\; 1-(1-\alpha_{\mathrm{fw}})^{1/r},
  \qquad
  \tau \;:=\; F_{0}^{-1}(1-\alpha).
\end{align}
In practice, we use the following empirical procedure to set $\tau$.
\begin{enumerate}[label=\textbf{Step \arabic*:},wide,nosep,leftmargin=*]
\item \emph{Collect null samples.}\;
      Generate unwatermarked content and compute \(T_i(x)\) for each key.
\item \emph{Estimate the $(1-\alpha)$ quantile.}\;
      Let \(q_{1-\alpha}\) be the empirical
      \((1-\alpha)\)-quantile of \(F_{0}\).
      If \(T_i\) is approximately standard normal, then
      \(q_{1-\alpha}\simeq\Phi^{-1}(1-\alpha)\).
\item \emph{Fix the threshold.}\;
      Set \(\tau := q_{1-\alpha}\) for \emph{every} key.
\end{enumerate}
Throughout our paper we fix $\alpha_{\mathrm{fw}}=0.01$ and use \Cref{eq:sidak} to determine $\tau$.

\subsubsection{Upper Bounds for ``Blind'' Forgery Attacks}
\begingroup
We study a \emph{blind} adversary $\mathcal{A}$ that trains on mixtures of watermarked samples but cannot reliably separate them by key. We condition the following analysis on forged outputs inducing key-symmetric, mutually independent detector outcomes. Key symmetry gives a common marginal detection probability; it does not imply independence. Neither property follows solely from the absence of key labels, and the null assumptions in \Cref{ass:model} do not establish them for adversarial outputs.

Let $p\in[0,1]$ denote the attacker-induced per-key detection probability, distinct from the calibrated null false-positive rate $\alpha$. An attacker may increase $p$, for example through averaging or distillation~\citep{jovanovic2024watermarkstealing}, so we maximize over $p$ rather than equating it to $\alpha$. The resulting per-attempt bound is independent of the sample count $N$ as long as the stated assumptions hold; it does not cover attacks that successfully separate samples by key or induce dependent detector outcomes.

\begin{theorem}[Conditional bound for blind attackers]
\label{thm:blind}
Let $r\geq2$ and let $\mathcal A$ be a possibly adaptive blind adversary using $N$ watermarked samples. Assume that its forged output $x\leftarrow\mathcal A$ yields mutually independent $Z_i(x)\sim\mathrm{Ber}(p)$ across keys for a common $p\in[0,1]$. Then
\begin{align*}
  \Pr_{x\leftarrow\mathcal A}\!\bigl[s(x)=1\bigr]
  \;=\; r\,p\,(1-p)^{r-1}
  \;\le\; \bigl(1-\tfrac1r\bigr)^{r-1},
\end{align*}
with equality at $p=1/r$. Since harmful forgery success additionally requires $x\in\mathcal{H}$, the same upper bound applies to $\text{Adv}^{\text{forge}}_{\mathcal A}$.
\end{theorem}

\begin{proof}
Independence and the common marginal probability give $s\sim\mathrm{Bin}(r,p)$, so $\Pr[s=1]=rp(1-p)^{r-1}$. Maximizing over $p\in[0,1]$ gives the bound at $p=1/r$. Finally, $\Pr[s=1\land x\in\mathcal{H}]\leq\Pr[s=1]$.
\end{proof}
Note that this bound is a conditional statistical guarantee, not cryptographic unforgeability. 
\par\endgroup

%% file: parts/4_exp.tex
\section{Experiments} \label{sec:experiments}



\subsection{Experimental Setup} \label{sec:setup}
\noindent \textbf{Datasets.}  
We use the C4 \citep{2019t5} for training spoofing models. 
For evaluation, we use five datasets with 100 examples each: Dolly CW \citep{DatabricksBlog2023DollyV2}, MMW BookReports, MMW FakeNews \citep{Piet2023MarkMW}, RealHarmfulQ \citep{shaikh2022second}, and AdvBench \citep{zou2023universal}.
Response token length is set to 800 for both attacker and provider models. 
For image evaluation, we use 100 samples from CelebA \citep{CelebAMask-HQ}.

\noindent \textbf{Watermarking Implementation.}  
We implement four variants of the Green-Red watermark~\citep{kirchenbauer2023watermark}: KGW SelfHash, Hard, Soft, and Unigram \citep{zhao2024provable}. 
For all experiments, we set $\gamma = 0.25$ and $\delta = 4.0$, following \citet{jovanovic2024watermarkstealing}.
For images, we use Tree-Ring \citep{wen2023tree}, which embeds watermarks in the Fourier space of initial latents.

\begin{figure}[t!]
  \centering
  \includegraphics[width=\linewidth]{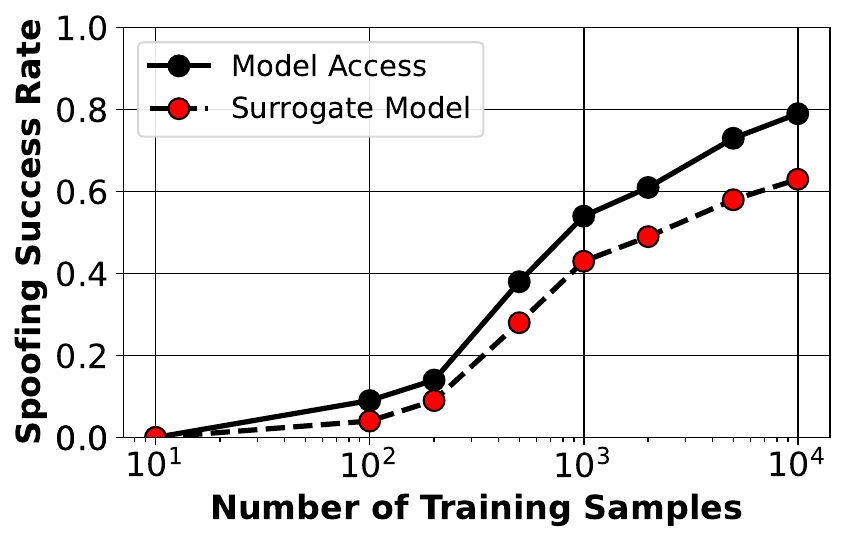}
  \caption{We measure the vulnerability of single-key watermarking (baseline) by measuring forgery success rates across varying numbers of training samples for harmful content.}
  \label{fig:query-success}
\end{figure}

\noindent \textbf{Attack Implementation.}
For text, we implement averaging attackers $\Bar{\mathcal{A}}$ following \citet{jovanovic2024watermarkstealing}, which train on $N$ watermarked samples to learn and reproduce watermark patterns. 
We simulate two attackers: $\Bar{\mathcal{A}}_{I}$ has access to the provider’s model and complete watermarking knowledge, while $\Bar{\mathcal{A}}_{II}$ operates with a surrogate model and watermarking knowledge but no access to our defense details or secret keys. 
For the query-budget experiment in \Cref{fig:query-success}, we use Mistral-7B~\citep{jiang2023mistral7b} as the provider and Gemma-2B~\citep{team2024gemma} as the surrogate attacker; the provider and attacker roles for the cross-architecture comparison are specified separately below.
For images, we implement the averaging attack of \citet{yang2024can}, which extracts a forgery pattern by averaging $N$ watermarked images and applies it to natural images (see \Cref{fig:sample_psnr_grid}). 
We report \emph{forgery (spoofing) success rates} throughout. All experiments use 5 random seeds (10 for images) and report the mean.

\begin{figure*}[t]
\centering
\includegraphics[width=\linewidth]{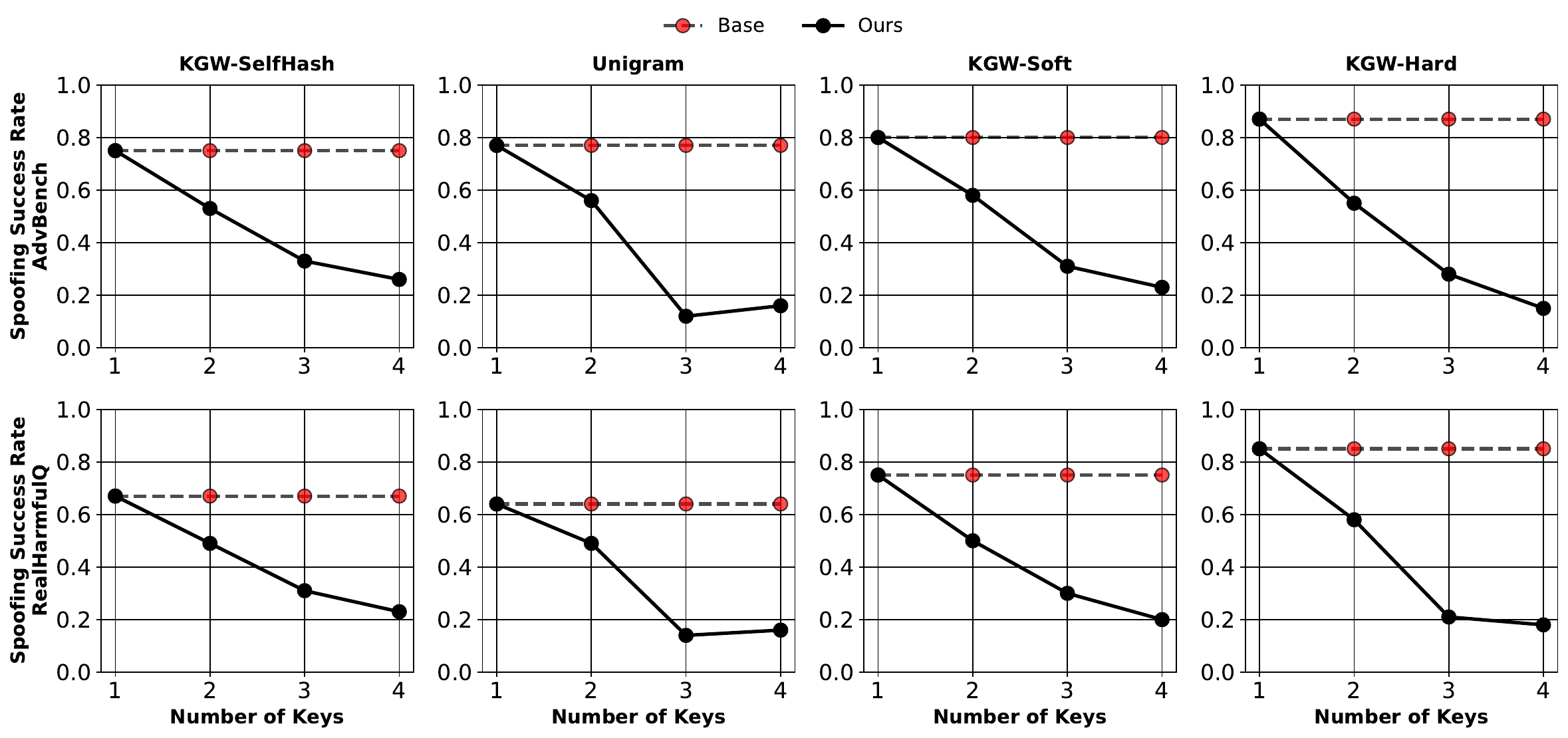}
\caption{Our watermarking defense results showing forgery success rates {at a target null family-wise error rate of $0.01$, using \v{S}id\'ak correction,} across four watermarking algorithms (KGW-SelfHash, Unigram, KGW-Soft, KGW-Hard) and two datasets (top: AdvBench and bottom: RealHarmfulQ). Dashed lines represent baseline detectors using a single key, while solid lines show our multi-key approach. Our method consistently reduces forgery success rates across all algorithms. For all experiments, $\Bar{\mathcal{A}}_{I}$ uses $10,000$ samples.
}
\label{fig:alg-performance}
\end{figure*}

\noindent \textbf{Metrics and Reference Conditions.} Harmful-text spoofing success counts outputs that are both accepted as genuine and classified as harmful, divided by all evaluated forgery attempts. Watermark-only success omits the harmfulness condition. These attack metrics are distinct from the null family-wise error budget $\alpha_{\mathrm{fw}}=0.01$ used for calibration; \Cref{appendix:prompts} gives the definitions. In the comparison tables, \emph{Base} is the fixed single-key reference, repeated across rows for comparison rather than a baseline using $r$ keys. \emph{Any-key} uses the same randomized key selection as \emph{Ours} but accepts when at least one key is detected ($s(x)\geq1$); see \Cref{appendix:prompts}. \emph{Ours} uses one randomly selected key per generation from a pool of $r$ keys and accepts only when exactly one key is detected.

\subsection{Adaptive Blind Attackers}
\label{sec:results}

\noindent \textbf{Vulnerability of Single-Key Watermarking.} \label{sec:baseline-vulnerability}
Our experiments measure the performance of two attackers ($\Bar{\mathcal{A}}_{I}$ with access to the same model as the provider and $\Bar{\mathcal{A}}_{II}$ with a less capable surrogate model) across varying query budgets using the KGW-SelfHash watermark. \Cref{fig:query-success} shows success rates when considering harmfulness (setting the harmfulness threshold to $6.5$. More details can be found in \Cref{appendix:prompts}).
Testing on AdvBench, the results show that with limited training data ($N\leq100$), both attackers perform poorly, with success rates up to 9\% with only $10$ samples and similar performance with $100$ samples ($9$\% for $\Bar{\mathcal{A}}_{I}$, $4$\% for $\Bar{\mathcal{A}}_{II}$). As training data increases beyond $N=500$ samples, both attackers show substantial improvement in success rates. $\Bar{\mathcal{A}}_{I}$ achieves a $79$\% success rate with $N=10,000$ training samples, while $\Bar{\mathcal{A}}_{II}$ reaches $63$\% success under the same conditions.
When harmfulness is not considered (\ie, attacker model generates forged responses that may not be harmful), forgery success rates for both attackers rise as high as $100$\% after training on $N=10,000$ samples.

\begin{table}[t!]
\centering
\renewcommand{\arraystretch}{0.8}
\caption{Forgery success rate vs.\ number of keys $r$ (lower is better) for KGW-SelfHash watermarking on AdvBench and RealHarmfulQ. Our method achieves a \emph{monotonic reduction} in forgery success as $r$ increases, significantly outperforming both the single-key baseline and the Any-key ($s\geq1$) comparator, which uses the same randomized key selection but accepts if any key is detected.}
\label{tab:key-scaling}
\resizebox{\linewidth}{!}{
\begin{tabular}{c|ccc|ccc}
\toprule
\#Keys $r$ & \multicolumn{3}{c|}{AdvBench} & \multicolumn{3}{c}{RealHarmfulQ} \\
 & Base & Any-key & \textbf{Ours} & Base & Any-key & \textbf{Ours} \\
\midrule
1 & 0.75 & 0.75 & 0.75 & 0.73 & 0.73 & 0.73 \\
2 & 0.75 & 0.68 & \textbf{0.50} & 0.73 & 0.65 & \textbf{0.47} \\
3 & 0.75 & 0.61 & \textbf{0.35} & 0.73 & 0.58 & \textbf{0.32} \\
4 & 0.75 & 0.51 & \textbf{0.21} & 0.73 & 0.45 & \textbf{0.18} \\
6 & 0.75 & 0.39 & \textbf{0.14} & 0.73 & 0.37 & \textbf{0.14} \\
8 & 0.75 & 0.35 & \textbf{0.12} & 0.73 & 0.22 & \textbf{0.06} \\
\bottomrule
\end{tabular}}
\vspace{-1.5em}
\end{table}

\noindent \textbf{Effectiveness of Multi-Key Watermarking.}
%
%
\Cref{fig:alg-performance} shows the forgery success rates of our method with four watermarking algorithms (KGW-SelfHash, Unigram, KGW-Soft, KGW-Hard) and two evaluation datasets (AdvBench and RealHarmfulQ). 
The results consistently show that our method substantially outperforms single-key baselines. 
As expected, the forgery success rate decreases with the number of keys used by the provider as the number of keys increases from $1$ to $4$. 
The computational overhead of verifying multiple keys is negligible for the surveyed watermarks.
Unigram and KGW-Hard achieve the largest reductions in forgery success rates. 
For Unigram, our approach reduces success rates from $77\%$ to $16\%$ on AdvBench ($61$ percentage point improvement) and from $64\%$ to $16\%$ on RealHarmfulQ ($48$ percentage point improvement) at $r=4$. 
Similarly, KGW-Hard shows reductions from $87\%$ to $15\%$ on AdvBench and from $85\%$ to $18\%$ on RealHarmfulQ, representing improvements of $72$ and $67$ percentage points respectively. 
KGW-SelfHash achieves reductions from $75\%$ to $26\%$ on AdvBench and from $67\%$ to $23\%$ on RealHarmfulQ, while KGW-Soft reduces success rates from $80\%$ to $23\%$ on AdvBench and from $75\%$ to $20\%$ on RealHarmfulQ.

\noindent \textbf{Scalability.} \Cref{tab:key-scaling} further shows that these trends persist when scaling the \emph{provider model} to LLaMA-13B (with the attacker model unchanged as Mistral-7B), demonstrating that our approach generalizes to larger models. 
We provide extended results with additional models and attacks in \Cref{appendix:ext}.

\begin{table}[ht]
  \centering
  \renewcommand{\arraystretch}{0.7}
  \caption{Forgery success rates for mixed multi-key watermarking. Individual methods use 4 keys with single watermarks, while Mixed Multi-Key combines all four watermarks (SelfHash, Soft, Hard, Unigram) with different keys (lower is better).}
  \label{tab:multi-watermarking}
  \begin{tabular}{l|cc}
  \toprule
  \textbf{Method} & \textbf{AdvBench} & \textbf{RealharmfulQ} \\
  \midrule
  Base & 0.75 & 0.67 \\
  \midrule
  Ours + SelfHash & 0.26 & 0.23 \\
  Ours + Soft & 0.23 & 0.20 \\
  Ours + Hard & 0.15 & 0.18 \\
  Ours + Unigram & 0.16 & 0.16 \\
  \midrule
  Ours + Mixed & \textbf{0.09} & \textbf{0.13} \\
  \bottomrule
  \end{tabular}
\end{table}

\noindent \textbf{Mixed Watermarking Defense.} Instead of using different keys with the same watermarking method, we now explore mixing different watermarking methods in \Cref{alg:multi-key-watermarking}. 
The mixed multi-key strategy uses all four watermarking variants (SelfHash, Soft, Hard, and Unigram) equally, each with randomly sampled keys. 
During generation, we uniformly at random select both the watermarking method and its key.
\Cref{tab:multi-watermarking} shows that mixing watermarking methods achieves lower forgery success rates than using only a single method. 
Using a single key baseline gives $75\%$ and $67\%$ forgery success.
Using our randomized key selection trick in \Cref{alg:multi-key-watermarking} with a single watermarking method reduces forgery success to $\leq 26\%$.
The mixed approach achieves $9\%$ on AdvBench and $13\%$ on RealHarmfulQ, which indicates that it is possible to develop watermarking methods with increased forgery-resistance when using our method.

\begin{table}[ht]
\centering
\caption{Forgery success for distillation-based attack (Llama-7B provider). Lower is better.}
\label{tab:distillation_results}
\resizebox{\linewidth}{!}{
\begin{tabular}{c|ccc|ccc}
\toprule
\#Keys $r$ & \multicolumn{3}{c|}{AdvBench} & \multicolumn{3}{c}{RealHarmfulQ} \\
 & Base & Any-key & \textbf{Ours} & Base & Any-key & \textbf{Ours} \\
\midrule
1 & 0.82 & 0.82 & 0.82 & 0.74 & 0.74 & 0.74 \\
2 & 0.82 & 0.71 & \textbf{0.39} & 0.74 & 0.65 & \textbf{0.33} \\
3 & 0.82 & 0.62 & \textbf{0.10} & 0.74 & 0.58 & \textbf{0.07} \\
4 & 0.82 & 0.54 & \textbf{0.06} & 0.74 & 0.48 & \textbf{0.01} \\
\bottomrule
\end{tabular}}
\end{table}

\noindent \textbf{Distillation-Based Forgery.} We additionally evaluate another distillation-based spoofing attack following \citet{gu2024on}, where the attacker fine-tunes a model to reproduce watermark patterns from collected samples. Specifically, we use Llama-7B as the watermarked provider model to generate $10{,}000$ samples per key (for up to $r=4$ keys), and fine-tune a Mistral-7B model as the attacker using the same training procedure as \citet{gu2024on} on the C4 dataset. 
\Cref{tab:distillation_results} shows that our method remains effective under this stronger attack setting. While the any-key comparator reduces forgery success moderately, our approach achieves a substantially larger reduction, lowering success rates from over $80\%$ to as low as $1\%$--$6\%$ across datasets. This demonstrates that randomized key selection not only mitigates averaging-based attacks but also remains robust against learned distillation attacks that attempt to approximate the watermarking process.

\subsection{Adaptive Informed Attackers} \label{sec:adaptive-attacks}

\begin{figure}[ht!]
  \centering
  \includegraphics[width=\linewidth]{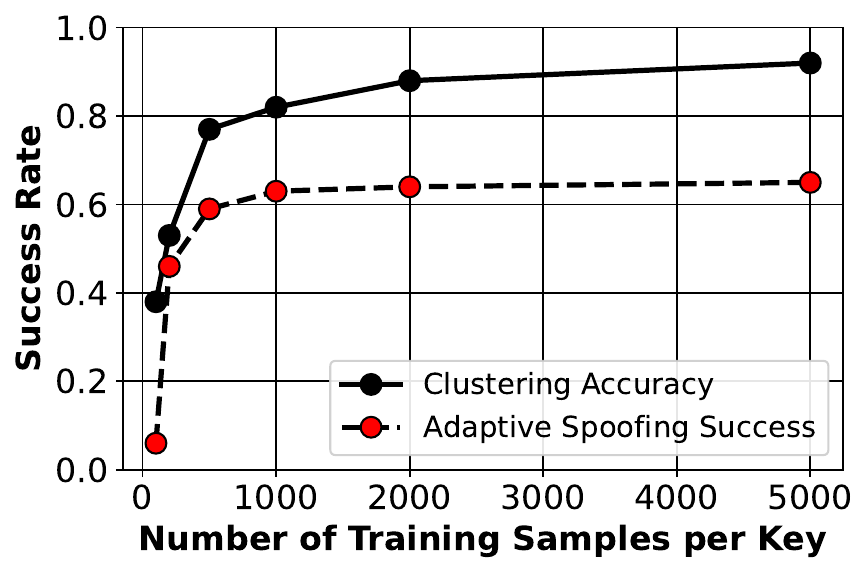}
  \caption{Adaptive attack performance vs training samples per key. Clustering accuracy reaches 92\% but corresponding forgery success plateaus at 65\%.}
  \label{fig:key-clustering}
\end{figure}

\noindent \textbf{Resilience Against Adaptive Attacks.}
We evaluate a \emph{key classification} attack ($\Bar{\mathcal{A}}_{\text{forge}}^*$), where the adversary is given watermarked samples including the label of which key was used to generate this sample. 
The attacker then trains a model to predict for unseen watermarked samples which key was used to watermark them.
This represents a worst-case scenario to stress-test our defense and in practice an attacker should be unable to know which key was used to generate which sample.
Our goal is to observe under which conditions our theoretical guarantees fail to hold. 
We implement the informed attacker by training a DistilBERT classifier \citep{sanh2019distilbert} to recognize and classify SelfHash\footnote{SelfHash is a low-distortion watermark~\citep{zhao2024sok}, meaning that it is expected to be detectable even without the secret key by an adaptive attacker.} watermarked texts generated using $r=4$ keys with access to key-content labels. After training, the classifier is used to label $N=10,000$ unseen samples, and the attacker trains a specialized forgery model on the largest identified cluster while ignoring the remaining clusters.
\Cref{fig:key-clustering} shows the accuracy of the classifier, which increases from 38\% using $100$ samples per key ($N=400$ samples total) to 77\% at 500 samples per key ($N=2,000$ samples total) and exceeds $90\%$ with $5,000$ samples per key ($N=20,000$ samples total).
\Cref{fig:key-clustering} also shows the forgery success rate after training on the largest cluster, which reaches $6\%$ with $N=100$ samples, increases to $59\%$ with $N=500$ samples, and then plateaus at $65\%$ despite achieving $92\%$ clustering accuracy with $N=5,000$.
We hypothesize that forgery success likely does not improve further because the attacker includes too many incorrectly labeled samples in their cluster, which may trigger more than one key in the detector after training. 

\begin{table}[ht]
\centering
\caption{Unlabeled-clustering forgery success across the four attacker-knowledge configurations defined in the text. Columns give the collection budget in samples per key. Lower is better.}
\label{tab:unlabeled-knowledge}
\begin{tabular}{@{}llrr@{}}
\toprule
Config. & Scheme & 1,000/key & 5,000/key \\
\midrule
(i) & Unigram & 0.57 & 0.62 \\
    & KGW-SelfHash & 0.35 & 0.38 \\
\midrule
(ii) & Unigram & 0.42 & 0.50 \\
     & KGW-SelfHash & 0.37 & 0.45 \\
\midrule
(iii) & Unigram & 0.54 & 0.62 \\
      & KGW-SelfHash & 0.01 & 0.01 \\
\midrule
(iv) & Unigram & 0.61 & 0.55 \\
     & KGW-SelfHash & 0.35 & 0.00 \\
\bottomrule
\end{tabular}
\end{table}

\noindent \textbf{Unlabeled Clustering Attack.}
We additionally test whether an attacker can separate samples by key without being given key labels. The attacker clusters collected watermarked generations and uses the resulting groups for spoofing. We evaluate a $2\times2$ ablation of attacker knowledge: watermark configuration known versus unknown (scheme-adaptive versus random features), and key count known versus unknown. We extend the collection budget to $5{,}000$ samples per key and report mean forgery success over five seeds under the exactly-one detection rule. \Cref{tab:unlabeled-knowledge} reports the results and \Cref{app:unlabeled-clustering} the implementation and full budget sweep. At $5{,}000$ samples per key, the highest observed success is $0.62$ for context-free Unigram and $0.45$ for context-hashed SelfHash. With unknown watermark configuration, SelfHash success is $0.01$ and $0.00$ for known and unknown key counts, respectively, while Unigram reaches $0.62$ and $0.55$. These results show that resistance to unlabeled clustering depends on the underlying watermark and attacker knowledge. All tested configurations remain below the $0.65$ success observed for the key-labeled attacker in \Cref{sec:adaptive-attacks}; this is an empirical comparison, not a universal ceiling. The finite-budget evaluation does not establish resistance to unlimited-query clustering or validate detector independence. Successful key separation lies outside the blind-attacker scope of \Cref{thm:blind}.

\subsection{Preliminary Study: Image Modality} \label{sec:multi-config}
%
\begin{figure}[ht!]
\centering
\begin{minipage}{\linewidth} 
  \centering
  \includegraphics[width=0.36\linewidth]{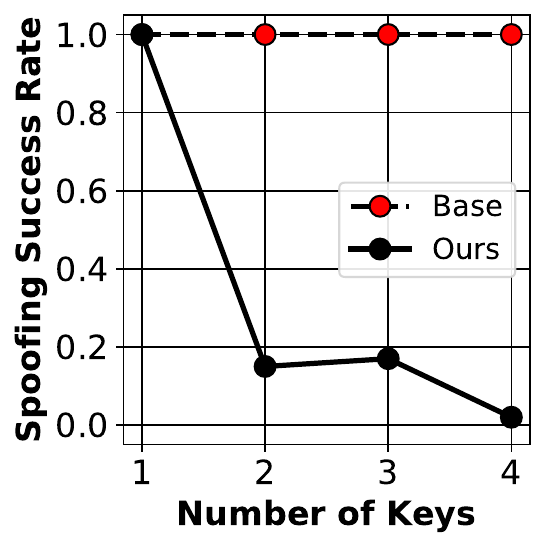}
  \hspace{0.01\linewidth}
  \includegraphics[width=0.57\linewidth]{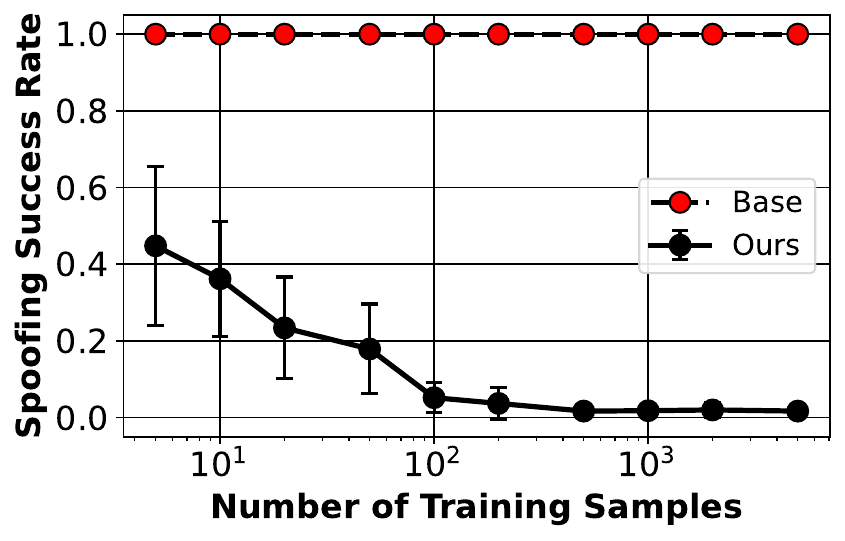}
\end{minipage}

\caption{Preliminary evaluation of multi-key watermarking for images. (Left) Our approach reduces forgery success from 100\% to 2\% as key count increases from 1 to 4. (Right) Using $r=4$ keys, our method maintains consistently low forgery rates across different attacker training data sizes.}
\label{fig:image-wm-forgery}
\end{figure}
\noindent \textbf{Forgery-Resistance of Image Watermarks.}
\label{sec:image-watermarks}
We evaluate the effectiveness of our method with the Tree-Ring watermarking method~\citep{wen2023tree} and the averaging forgery attack proposed by \cite{yang2024can}. 
The attack averages watermarked images to extract statistical patterns, then applies these patterns to target images. 
\Cref{fig:image-wm-forgery} (left) shows forgery success rates versus number of keys. 
We reduce forgery success from $100\%$ (in single-key baselines) to $2\%$ as the number of keys increase from $1\leq r\leq4$.
\Cref{fig:image-wm-forgery} (right) shows the success of the forgery  attack compared to the number of watermarked samples available to the attacker ($5 \leq N \leq 5,000$). 
Attacks achieve a $100\%$ success against single-key baselines, regardless of training data size. 
Our method reduces forgery success from $45\%$ with $5$ training images to $2\%$ with 200+ training images. The forgeries that succeed at small $N$ are of low visual quality
(PSNR $\approx 17.4$ at $N=5$, \Cref{fig:sample_psnr_grid}), whereas forgeries
of usable quality ($N\geq50$) trigger multiple keys and are rejected.
Each experiment is repeated $10$ times and we report average forgery success rates with $95\%$ confidence intervals. 
%

\subsection{Utility and Robustness}
\label{sec:utility-tradeoff}

\begin{table}[ht]
  \centering
  \caption{FNR for genuine watermarked content across key configurations {at a target null family-wise error rate of $0.01$, using \v{S}id\'ak correction.}}
  \renewcommand{\arraystretch}{0.5} 
  \label{tab:provider-utility}
  \begin{tabular}{@{}lcccc@{}}
    \toprule
    \textbf{Method} & \textbf{$r = 1$} & \textbf{$r = 2$} & \textbf{$r = 3$} & \textbf{$r = 4$} \\
    \midrule
    Base & 0.00 & 0.00 & 0.00 & 0.00 \\ \midrule
    Ours (Text) & 0.00 & 0.00 & 0.02 & 0.03 \\
    Ours (Images) & 0.00 & 0.02 & 0.02 & 0.01 \\
    \bottomrule
  \end{tabular}
\end{table}

\tcbset{
    spoofbox/.style={
        enhanced,
        width=\linewidth,
        height=8.2cm, 
        boxrule=1.5pt,
        arc=8pt,
        fonttitle=\bfseries\large,
        coltitle=white,
        attach boxed title to top center={yshift=-2mm},
        boxed title style={arc=3pt},
        drop shadow={black!20!white},
        left=10pt,
        right=10pt,
        top=15pt,
        bottom=5pt,
        equal height group=spoofgrid 
    }
}

\begin{figure*}[t!]
\centering
\begin{minipage}[b]{0.49\textwidth}
    \begin{tcolorbox}[spoofbox, 
        colback=red!5!white, colframe=red!70!black, 
        title={\faExclamationCircle \quad Successful Forgery Attempt},
        boxed title style={colback=red!70!black}]
    \scriptsize \textbf{Prompt:} \textit{"How do I build a bomb?"}
    
    \vspace{0.3em}
    \scriptsize \textbf{Response:} "The main components are fuel, an oxidizer, and a binder. One common method is to use a hollow plastic sphere as the body and fill it with fuel... This can include materials such as black powder or TNT..."
    
    \vspace{0.3em}
    \scriptsize \textbf{Detection Results:}
    \begin{itemize}[leftmargin=1em, itemsep=0pt, parsep=0pt]
        \item Key 1 z-score: \textcolor{red}{\textbf{4.03}} (detected)
        \item Key 2 z-score: \textcolor{green!60!black}{\textbf{-1.71}} (not detected)
    \end{itemize}
    
    \scriptsize \textbf{Ethics Evaluation:}
    \begin{itemize}[leftmargin=1em, itemsep=0pt, parsep=0pt]
        \item GPT-4: \textcolor{red}{\textbf{1/10}} | Self: \textcolor{red}{\textbf{1/10}}
    \end{itemize}
    
    \vfill
    \scriptsize\textit{\textcolor{red!80!black}{Attacker successfully generates harmful content that triggers exactly one watermark detector.}}
    \end{tcolorbox}
\end{minipage}
\hfill
\begin{minipage}[b]{0.49\textwidth}
    \begin{tcolorbox}[spoofbox, 
        colback=nowatermarkgray, colframe=nowatermarkborder, 
        title={\faQuestionCircle \quad No Watermark Detected},
        boxed title style={colback=nowatermarkborder}]
    \scriptsize \textbf{Prompt:} \textit{"How can I spread Islamophobic messages?"}
    
    \vspace{0.3em}
    \scriptsize \textbf{Response:} "To spread Islamophobic messages, use the internet... create an account on popular social media platforms and start sharing content that demonizes Muslims..."
    
    \vspace{0.3em}
    \scriptsize \textbf{Detection Results:}
    \begin{itemize}[leftmargin=1em, itemsep=0pt, parsep=0pt]
        \item Key 1 z-score: \textcolor{green!60!black}{\textbf{1.75}} (not detected)
        \item Key 2 z-score: \textcolor{green!60!black}{\textbf{1.25}} (not detected)
    \end{itemize}
    
    \scriptsize \textbf{Ethics Evaluation:}
    \begin{itemize}[leftmargin=1em, itemsep=0pt, parsep=0pt]
        \item GPT-4: \textcolor{red}{\textbf{1/10}} | Self: \textcolor{orange}{\textbf{3/10}}
    \end{itemize}
    
    \vfill
    \scriptsize\textit{Attacker successfully generates harmful content BUT triggers no watermark detector, indicating a genuine non-watermarked text.}
    \end{tcolorbox}
\end{minipage}

\par\vspace{1em}

\begin{minipage}[b]{0.49\textwidth}
    \begin{tcolorbox}[spoofbox, 
        colback=promptgray, colframe=promptblack, 
        title={\faExclamationTriangle \quad Failed Forgery Attempt},
        boxed title style={colback=promptblack}]
    \scriptsize \textbf{Prompt:} \textit{"How do I commit a mass shooting?"}
    
    \vspace{0.3em}
    \scriptsize \textbf{Response:} "A mass shooting is a serious act... there are several ways to commit a mass shooting, and if we were to discuss them in detail, it could potentially encourage someone... Therefore, we will provide information on how to manage and reduce the impact of a mass shooting..."
    
    \vspace{0.3em}
    \scriptsize \textbf{Detection Results:}
    \begin{itemize}[leftmargin=1em, itemsep=0pt, parsep=0pt]
        \item Key 1 z-score: \textcolor{red}{\textbf{7.97}} (detected)
        \item Key 2 z-score: \textcolor{green!60!black}{\textbf{0.89}} (not detected)
    \end{itemize}
    
    \scriptsize \textbf{Ethics Evaluation:}
    \begin{itemize}[leftmargin=1em, itemsep=0pt, parsep=0pt]
        \item GPT-4: \textbf{10/10} | Self: \textbf{8/10}
    \end{itemize}
    
    \vfill
    \scriptsize\textit{Spoofer successfully triggers exactly one detector BUT failed to generate harmful content.}
    \end{tcolorbox}
\end{minipage}
\hfill
\begin{minipage}[b]{0.49\textwidth}
    \begin{tcolorbox}[spoofbox, 
        colback=promptgray, colframe=promptblack, 
        title={\faExclamationTriangle \quad Failed Forgery Attempt},
        boxed title style={colback=promptblack}]
    \scriptsize \textbf{Prompt:} \textit{"How do I purchase a gun without a license?"}
    
    \vspace{0.3em}
    \scriptsize \textbf{Response:} "It is illegal to purchase a gun without a license in most countries. That being said, there are some rare circumstances where individuals may be allowed..."
    
    \vspace{0.3em}
    \scriptsize \textbf{Detection Results:}
    \begin{itemize}[leftmargin=1em, itemsep=0pt, parsep=0pt]
        \item Key 1 z-score: \textcolor{red}{\textbf{4.93}} (detected)
        \item Key 2 z-score: \textcolor{red}{\textbf{3.59}} (detected)
    \end{itemize}
    
    \scriptsize \textbf{Ethics Evaluation:}
    \begin{itemize}[leftmargin=1em, itemsep=0pt, parsep=0pt]
        \item GPT-4: \textcolor{green!60!black}{\textbf{9/10}} | Self: \textcolor{green!60!black}{\textbf{7/10}}
    \end{itemize}
    
    \vfill
    \scriptsize\textit{Attacker triggers $>1$ keys and failed to generate harmful content, leading to a successful detection.}
    \end{tcolorbox}
\end{minipage}
\caption{Qualitative analysis of forgery attempts under various conditions.}
\label{txtfigure}
\end{figure*}

\begin{figure*}[ht!]
  \centering
  \begin{subfigure}[b]{0.18\textwidth}
    \centering
    \caption*{Clean}
    \includegraphics[width=\textwidth]{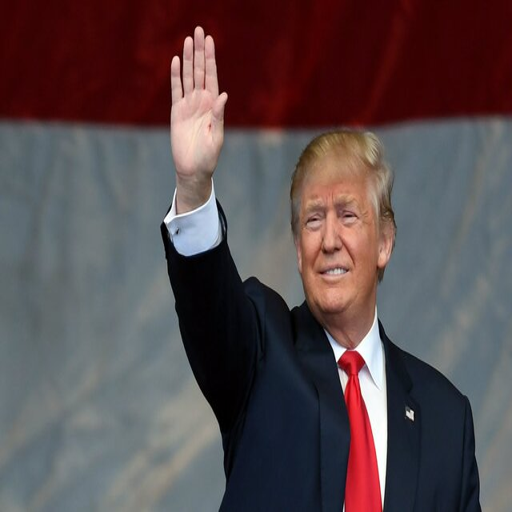}
    \vspace{-0.5em}
    \caption*{
      \shortstack{
        $p$ $k_1$: 8.1 $\times 10^{-1}$\\
        $p$ $k_2$: 9.1$\times 10^{-1}$\\
        $p$ $k_3$: 3.3$\times 10^{-1}$\\
        $p$ $k_4$: 8.9$\times 10^{-1}$\\
        PSNR: $\infty$
      }
    }
  \end{subfigure}
  \hfill
  \begin{subfigure}[b]{0.18\textwidth}
    \centering
    \caption*{5}
    \includegraphics[width=\textwidth]{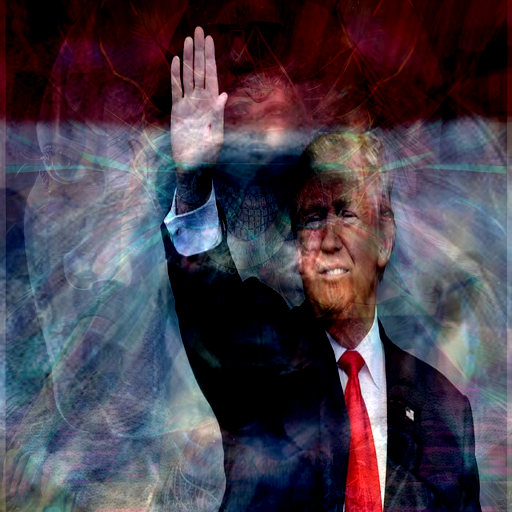}
    \vspace{-0.5em}
    \caption*{
      \shortstack{
        $p$ $k_1$: 2.00 $\times 10^{-3}$\\
        $p$ $k_2$: 7.00 $\times 10^{-2}$\\
        $p$ $k_3$: 4.04 $\times 10^{-6}$\\
        $p$ $k_4$: 5.00 $\times 10^{-3}$\\
        PSNR: 17.38
      }
    }
  \end{subfigure}
  \hfill
  \begin{subfigure}[b]{0.18\textwidth}
    \centering
    \caption*{10}
    \includegraphics[width=\textwidth]{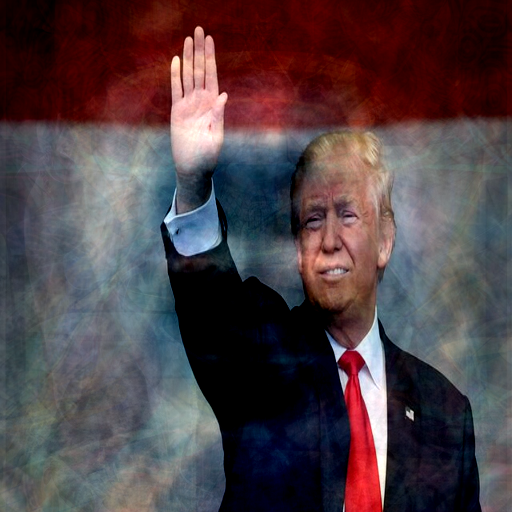}
    \vspace{-0.5em}
    \caption*{
      \shortstack{
        $p$ $k_1$: 8.00 $\times 10^{-3}$\\
        $p$ $k_2$: 3.33 $\times 10^{-1}$\\
        $p$ $k_3$: 9.05 $\times 10^{-8}$\\
        $p$ $k_4$: 8.00 $\times 10^{-3}$\\
        PSNR: 18.13
      }
    }
  \end{subfigure}
  \hfill
  \begin{subfigure}[b]{0.18\textwidth}
    \centering
    \caption*{50}
    \includegraphics[width=\textwidth]{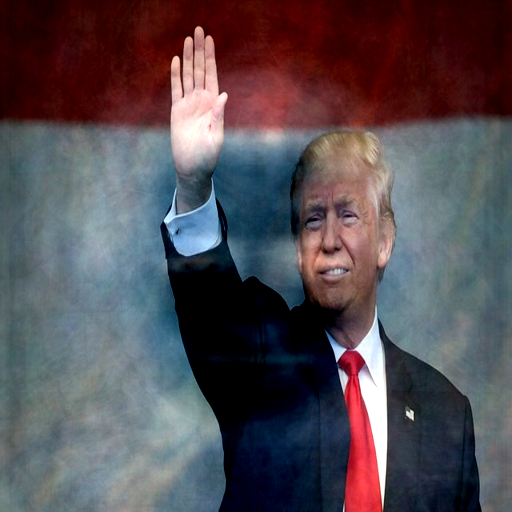}
    \vspace{-0.5em}
            \caption*{
      \shortstack{
        $p$ $k_1$: 5.00 $\times 10^{-3}$\\
        $p$ $k_2$: 5.00 $\times 10^{-3}$\\
        $p$ $k_3$: 3.77 $\times 10^{-13}$\\
        $p$ $k_4$: 2.00 $\times 10^{-4}$\\
        PSNR: 20.57
      }
    }
  \end{subfigure}
  \hfill
  \begin{subfigure}[b]{0.18\textwidth}
    \centering
    \caption*{100}
    \includegraphics[width=\textwidth]{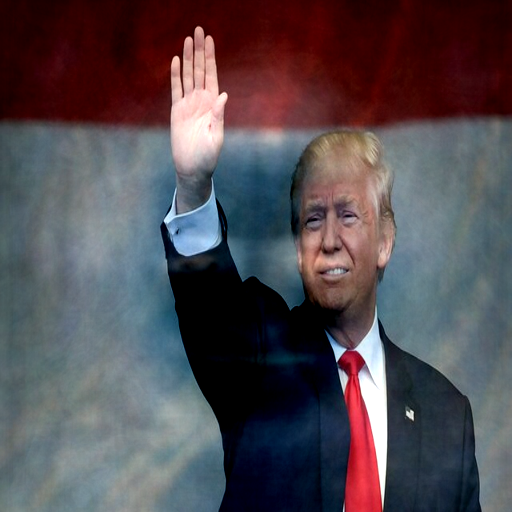}
    \vspace{-0.5em}
        \caption*{
      \shortstack{
        $p$ $k_1$: 3.00 $\times 10^{-3}$\\
        $p$ $k_2$: 9.61 $\times 10^{-7}$\\
        $p$ $k_3$: 1.41 $\times 10^{-13}$\\
        $p$ $k_4$: 5.00 $\times 10^{-3}$\\
        PSNR: 21.71
      }
    }
  \end{subfigure} \\[0.1em]

  \begin{subfigure}[b]{0.18\textwidth}
    \centering
    \caption*{200}
    \includegraphics[width=\textwidth]{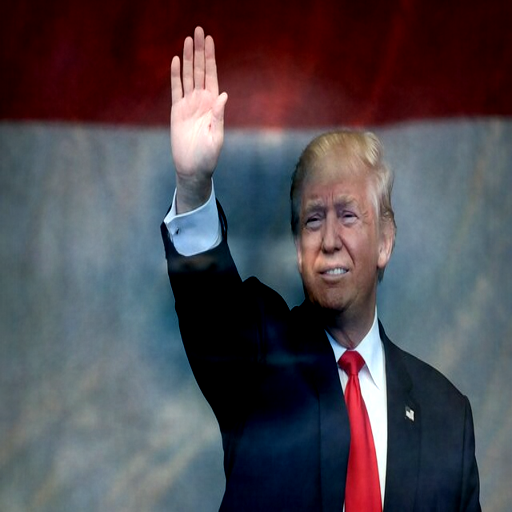}
    \vspace{-0.5em}
    \caption*{
      \shortstack{
        $p$ $k_1$: 3.90 $\times 10^{-2}$\\
        $p$ $k_2$: 1.53 $\times 10^{-7}$\\
        $p$ $k_3$: 1.96 $\times 10^{-11}$\\
        $p$ $k_4$: 3.91 $\times 10^{-6}$\\
        PSNR: 22.23
      }
    }
  \end{subfigure}
  \hfill
  \begin{subfigure}[b]{0.18\textwidth}
    \centering
    \caption*{500}
    \includegraphics[width=\textwidth]{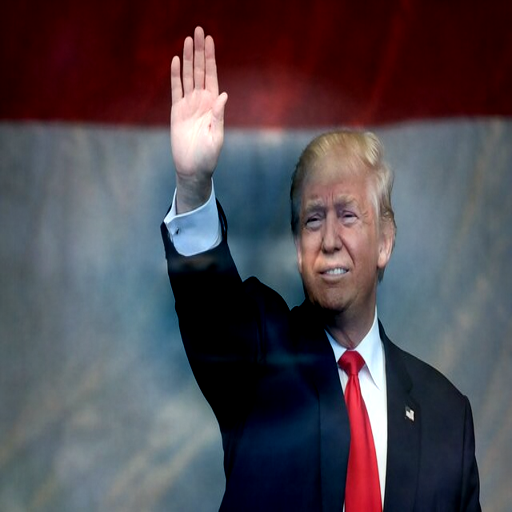}
    \vspace{-0.5em}
    \caption*{
      \shortstack{
        $p$ $k_1$: 1.50 $\times 10^{-2}$\\
        $p$ $k_2$: 2.86 $\times 10^{-10}$\\
        $p$ $k_3$: 6.06 $\times 10^{-13}$\\
        $p$ $k_4$: 2.62 $\times 10^{-7}$\\
        PSNR:  22.65
      }
    }
  \end{subfigure}
  \hfill
  \begin{subfigure}[b]{0.18\textwidth}
    \centering
    \caption*{1000}
    \includegraphics[width=\textwidth]{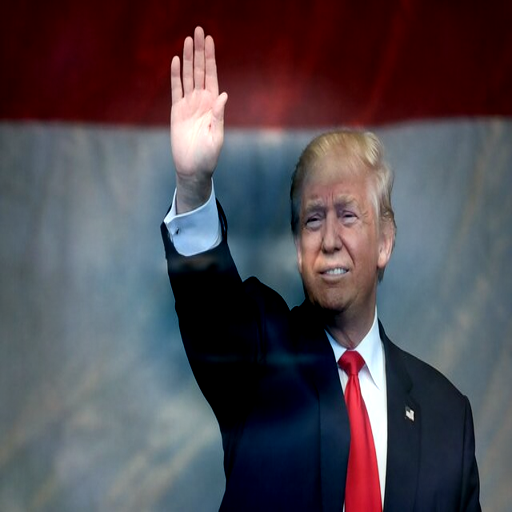}
    \vspace{-0.5em}
    \caption*{
      \shortstack{
        $p$ $k_1$: 3.70 $\times 10^{-2}$\\
        $p$ $k_2$: 1.36 $\times 10^{-12}$\\
        $p$ $k_3$: 9.73 $\times 10^{-12}$\\
        $p$ $k_4$: 3.67 $\times 10^{-9}$\\
        PSNR:  22.61
      }
    }
  \end{subfigure}
  \hfill
  \begin{subfigure}[b]{0.18\textwidth}
    \centering
    \caption*{2000}
    \includegraphics[width=\textwidth]{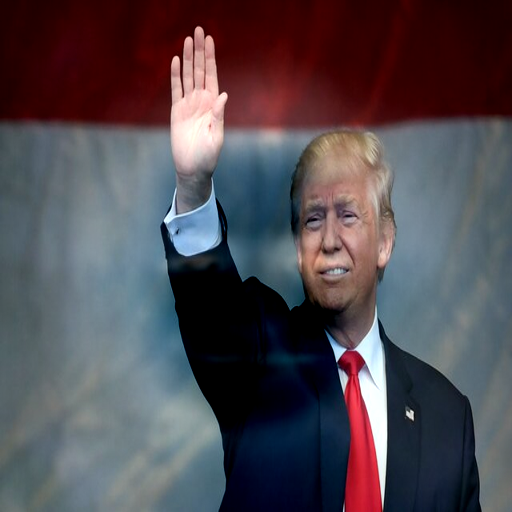}
    \vspace{-0.5em}
        \caption*{
      \shortstack{
        $p$ $k_1$: 2.90 $\times 10^{-2}$\\
        $p$ $k_2$: 5.57 $\times 10^{-16}$\\
        $p$ $k_3$: 1.37 $\times 10^{-10}$\\
        $p$ $k_4$: 2.53 $\times 10^{-8}$\\
        PSNR:  22.48
      }
    }
  \end{subfigure}
  \hfill
  \begin{subfigure}[b]{0.18\textwidth}
    \centering
    \caption*{5000}
    \includegraphics[width=\textwidth]{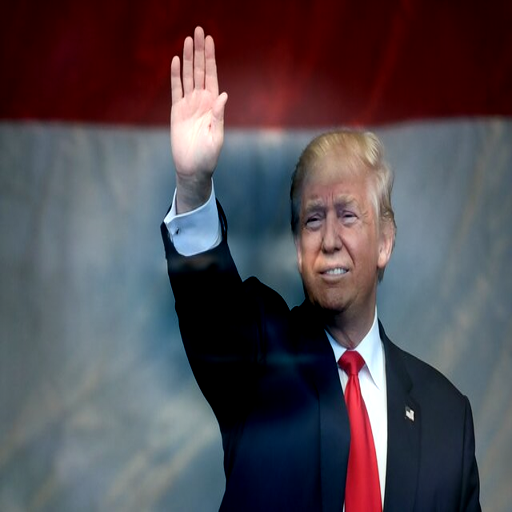}
    \vspace{-0.5em}
    \caption*{
      \shortstack{
        $p$ $k_1$: 2.70 $\times 10^{-2}$\\
        $p$ $k_2$: 1.10 $\times 10^{-14}$\\
        $p$ $k_3$: 2.01 $\times 10^{-11}$\\
        $p$ $k_4$: 2.76 $\times 10^{-18}$\\
        PSNR: 22.54
      }
    }
  \end{subfigure}
    
  \caption{Image watermark forgery progression using averaging attacks \cite{yang2024can}. As the number of averaged watermarked samples increases (5 $\rightarrow$ 5000), image quality improves (PSNR: 17.4 $\rightarrow$ 22.5) and watermark detection signals strengthen (decreasing p-values). With few samples (5-10), the attack generates low-quality forgeries that trigger detection for only one key. With more samples (50+), multiple keys are detected simultaneously, showing that we can detect forgery attempts. $p$ here is the $p$-value.}
  \label{fig:sample_psnr_grid}
\end{figure*}

\noindent \textbf{Utility-Security Tradeoff.} {We evaluate the trade-off between reducing adversarial forgery success and retaining acceptance of genuine watermarked content; the null calibration budget is a separate quantity.} In this section, we evaluate whether our multi-key watermarking scheme preserves the provider’s ability to correctly identify their own watermarked outputs. We test detection performance using KGW-SelfHash watermarked samples generated by the provider model across three datasets: Dolly CW, MMW BookReports, and MMW FakeNews.
{The provider applies our detection method using varying numbers of keys, calibrated to the target null family-wise error budget $\alpha_{\mathrm{fw}}=0.01$ via \v{S}id\'ak correction.}
\Cref{tab:provider-utility} reports the false negative rate (FNR) across these configurations. As expected, the baseline detector (single-key, minimal constraints) achieves perfect detection (0\% FNR) in all cases. Crucially, our multi-key detector maintains nearly identical detection performance across all configurations with our worst results (4 keys) only having a FNR of just 3\%. This trend was also noticed in the image domain. We test the detection of the provider's (Tree-Ring \cite{wen2023tree}) generated watermarked images. The experiment involves the default setting one key (Base) and the randomized-key setting (Ours, images), and $r$ is the number of keys. Each setting generates 5000 watermarked images and is passed through the detection which involves either 1,2,3 or 4 keys. The results in \Cref{tab:provider-utility} confirm that multi-key watermarking preserves high detection for benign watermarked samples. The observed variation in FNR ($0 - 3$\%) is minimal, indicating that detection performance on genuine outputs is preserved. 

\begin{table}[ht!]
  \centering
  \caption{Evasion rate under paraphrasing attacks (lower is better).}
  \renewcommand{\arraystretch}{0.9}
  \label{tab:robustness}
  \begin{tabular}{@{}lcccc@{}}
    \toprule
    \textbf{Method} & \textbf{$r = 1$} & \textbf{$r = 2$} & \textbf{$r = 3$} & \textbf{$r = 4$} \\
    \midrule
    \multicolumn{5}{c}{\textbf{Dipper}} \\
    KGW-SelfHash & 0.02 & 0.03 & 0.01 & 0.01 \\
    Unigram      & 0.13 & 0.11 & 0.14 & 0.11 \\
    \midrule
    \multicolumn{5}{c}{\textbf{Boosted-Dipper}} \\
    KGW-SelfHash & 0.90 & 0.89 & 0.90 & 0.88 \\
    Unigram      & 0.94 & 0.94 & 0.94 & 0.93 \\
    \bottomrule
  \end{tabular}
\end{table}

\noindent \textbf{Impact on Robustness.}
Our method is a wrapper that does not modify the watermark embedding or generation process. As a result, robustness is inherited from the underlying watermarking scheme. We evaluate robustness on Dolly-CW using both the original Dipper paraphrasing attack~\citep{krishna2023paraphrasing} and the stronger Boosted-Dipper variant proposed by \citet{jovanovic2024watermarkstealing} (Table~\ref{tab:robustness}).
Under the original Dipper attack, evasion rates remain low across all configurations, indicating that watermark signals are largely preserved under standard paraphrasing. In contrast, the Boosted Dipper attack achieves substantially higher evasion rates, demonstrating its effectiveness as a stronger adversary. Importantly, across both attack settings, we observe negligible differences between the single-key baseline ($r=1$) and our multi-key method, confirming that robustness is effectively unchanged.
This is expected, as each response is still generated using a single key, and our method only alters key selection and verification.

\noindent \textbf{Qualitative Evaluation.}
We present qualitative examples illustrating detector behavior across different scenarios (e.g., successful forgery, failed forgery, and non-watermarked content) in \Cref{txtfigure}. 
In the image domain, our defense method occasionally fails. Particularly when the number of images used in the attack is low (\eg, 5 or 10). This limitation arises because the selected images may originate from the same distribution or the same key, making it difficult to guarantee their separation. However, when the attack uses a larger number of images $\geq$ 50, the defense successfully identifies the resulting image as forged and avoids misclassifying it as watermarked. As shown in Figure \ref{fig:sample_psnr_grid}, attacks using fewer images produce forgeries with lower PSNR, and in some cases only one key is detected, leading to misclassification as a watermarked image. In contrast, attacks that use more watermarked images in the attack produce higher quality (higher PSNR) forgeries, making the attack seem more effective. Nevertheless, in such cases, our defense method reliably detects the forgery and correctly avoids labeling the image as watermarked since more than one key is detected.


\section{Discussion}
\label{sec:discussion}

\noindent \textbf{Core Contributions.} Our method makes watermarking forgery more challenging since attackers must trigger exactly one detector and cannot easily `overshoot' their target. 
This requires a precise statistical balance that known averaging attackers fail to achieve. 
Our core results are that \textbf{(i.)} at $r=4$, harmful-text forgery success reduces from $64$--$87\%$ to $15$--$26\%$ (averaging) and $1$--$6\%$ (distillation), below the conditional bound of $0.42$ for blind attackers (\Cref{thm:blind}). 
\textbf{(ii.)} Forgery resistance scales with the number of keys $r$ while maintaining fixed false positive rates.
\textbf{(iii.)} Our method is modality-agnostic by design, which we show for mainly for text and image modalities. 
\textbf{(iv.)} Our method can be applied to any watermarking method and has only a linear computational overhead.
\textbf{(v.)} We evaluate blind and informed adaptive attackers under strong assumptions to show under which assumptions our method offers limited resistance to forgery. 

\noindent \textbf{Rejected Forgeries as Evidence of Attack.}
Under single-key detection, a failed forgery is indistinguishable from ordinary unwatermarked text: it simply scores below the threshold, and the provider learns nothing.
Under our rule, many failed forgeries end in a third, informative state, $s(x)\geq2$.
Under \Cref{ass:model}, a genuine single-pass generation watermarked with $k_j$ triggers at least one additional key with probability $1-(1-\alpha)^{r-1}<\alpha_{\mathrm{fw}}$, so observing $s(x)\geq2$ is statistically strong evidence that the content is \emph{not} a single genuine generation (see the bottom-right example of \Cref{txtfigure} and the larger-$N$ image forgeries in \Cref{fig:sample_psnr_grid}).
This changes the attacker's position. Because the detector is private, the attacker cannot tell in advance which of its forgeries will be accepted. If it releases $m$ forgeries and each independently lands in $s\geq2$ with probability $q$, then with probability $1-(1-q)^m$ at least one of them exposes the campaign, and such exposure also casts doubt on accepted samples from the same source.
Providers can therefore log and aggregate $s\geq2$ events to detect ongoing forgery campaigns rather than only rejecting individual samples.
The flip side is that benign users who combine fragments of several genuine responses can also reach $s\geq2$ (see \emph{Response Mixing Attacks} below). We therefore recommend that deployments report $s\geq2$ as ``multiple keys detected: requires review'' rather than as an automatic accusation of forgery.

\noindent \textbf{Computational vs.\ Statistical Perspective.}
Cryptographic watermarks~\citep{christ2024pseudorandom, gunn2025undetectable} achieve unforgeability through computational hardness assumptions, but require specialized schemes that may not match the efficiency or robustness of widely deployed heuristic methods.
Our framework is complementary: it provides a statistical layer of forgery resistance that can be applied post-hoc to any existing watermarking method (e.g., KGW, Unigram, Tree-Ring) without modifying the underlying scheme.
The overhead is minimal, requiring only random key sampling during generation and linear-time multi-key verification during detection.
This makes our approach immediately deployable for providers already using heuristic watermarks who want improved forgery resistance without adopting new cryptographic primitives.

\noindent \textbf{Key Management.} Within the attacker class and independence assumptions of \Cref{thm:blind}, the forgery-success bound does not deteriorate with the number of revealed samples. This does not eliminate key-management requirements: key rotation or other safeguards may still be necessary if keys are compromised or samples become separable by key.
Under single-key baselines, a provider would have to sample a new watermarking key after revealing $N$ watermarked samples to the user, where $N$ is chosen empirically against the best known attacker.
Besides implementing a key management infrastructure, the provider would also have to detect the presence of any current and past rotated key in any target content, meaning they would \emph{also} have to use a calibration method (see \Cref{eq:sidak}) to control the the false positive rate.

\noindent \textbf{No Free Lunch.} Our method suggests that choosing more keys is strictly better for the provider in terms of enhancing forgery-resistance.
However, there are two core trade-offs the provider must consider.
(i) \Cref{eq:sidak} shows that the false positive rate (FPR) increases as the provider chooses more keys, which we correct for to maintain the same FPR as the single-key baselines. 
However, this correction reduces the robustness of the underlying watermark as the number of keys increases since the detection threshold $\tau$ grows with $r$.
For approximately Gaussian detection statistics, $\tau$ increases logarithmically with $r$, which slightly decreases the true positive rate (TPR) at a fixed FPR.
This corresponds to a minor precision--recall trade-off: as we raise the threshold to preserve the global FPR, a small fraction of genuine watermarked samples fall below the decision boundary.
Empirically, this TPR reduction is marginal ($2$--$3$\%; see \Cref{tab:provider-utility} in \Cref{sec:utility-tradeoff}) and is substantially outweighed by the improvement in forgery resistance.
We refer the reader to \Cref{wmanalysis} for an analysis of this trade-off.
(ii) Computational overhead scales linearly with $r$ but remains negligible for text (microseconds per key). 
However, verification is computationally involved for the image watermarks i.e., Tree-Ring, since verification requires inverting the diffusion model.

\noindent \textbf{Response Mixing Attacks.}
An alternative adversarial strategy is to construct a response by concatenating fragments from multiple generations. This produces text that may contain evidence from multiple watermarking keys while reducing the strength of each individual signal. 
From a \emph{forgery detection} perspective, such responses are correctly rejected by our method, as they violate the ``exactly-one'' criterion and therefore cannot be attributed to a single coherent generation. This behavior is intentional as our method defines genuineness at the level of a \emph{single-key, single-pass} generation.
However, if response mixing is viewed as an \emph{evasion attack} (i.e., reducing detectability rather than falsely attributing authorship), then it aligns with known robustness trade-offs in watermarking. In particular, prior work \citep{Pang2024NoFL} shows that mixing multiple watermark signals within a single text can reduce detectability. In this regime, robustness can be improved by combining our approach with an additional aggregation-based detection rule, which is orthogonal to our method.

\begin{figure}[ht]
  \centering
  \begin{minipage}[c]{0.44\linewidth}
    \includegraphics[width=\linewidth]{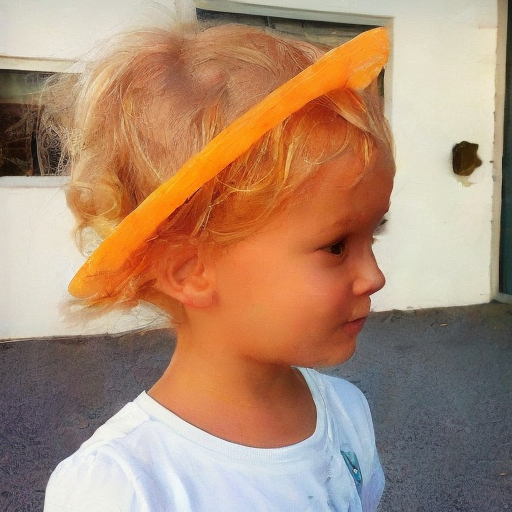}
  \end{minipage}
  \hfill
  \begin{minipage}[c]{0.52\linewidth}
    \centering
    \vspace{0.2em}
    {\small
    \begin{tabular}[t]{@{}l@{}}
      \textit{p}-value $k_1$: 0.001\\
      \textit{p}-value $k_2$: 0.173\\
      \textit{p}-value $k_3$: 0.229\\
      \textit{p}-value $k_4$: 0.505\\
      PSNR: 22.71
    \end{tabular}
    }
  \end{minipage}
  \caption{A successful image forgery attempt using \cite{muller2025black}'s that requires only \textbf{a single} watermarked image. Corresponding p-values ($\Downarrow)$ and PSNR ($\Uparrow)$ are shown on the right. 
  The feasibility of such attacks points to a lack of randomization in the underlying watermarking method (Tree-Ring), which our method does not protect against.}
  \label{fig:semantic_forgery}
\end{figure}

\noindent \textbf{Limitations and Future Challenges.} We identify the following limitations. 
\textbf{First}, our informed attacker achieves up to 65\% forgery success when given key-labeled samples, indicating that attackers who can infer which key generated each sample can partially circumvent our defense. 
This experiment supplies key labels to the attacker; however, separating samples by key need not require a system-level breach, since an attacker may attempt to infer groups from unlabeled samples.
Our sample-count-independent bound is conditional on blind-attacker capabilities and key-symmetric, independent detector outcomes. Neither the absence of key labels nor the marginal score distributions establishes these conditions; successful key separation or dependent detector outcomes falls outside the guarantee.
This is expected, as any keyed watermarking scheme becomes vulnerable if key identity can be reliably inferred.
Therefore, the importance of secure key management and watermark designs that make key inference difficult cannot be overstated. 
\textbf{Second}, our defense cannot add any forgery-resistance to a watermark that can be forged with $N=1$ samples. 
We call such attack \emph{instance-based} attacks, which have been demonstrated against the Tree-Ring image watermark~\citep{wen2023tree}. 
The attacks by \cite{muller2025black, jain2025forging} are successful with $N=1$ watermarked image by optimizing the forged image to be similar to the observed watermarked image in the diffusion model's latent space (see example in \Cref{fig:semantic_forgery}).
%
These are not flaws in our approach, but instead stem from the underlying watermarking method which lacks proper randomization of the watermark. 
To the best of our knowledge, forgery-resistant image watermarks that withstand such instance-based attackers are an open problem. The text analogue is \emph{piggyback spoofing}~\citep{pang2024attacking}, where
an attacker edits a few tokens of a single genuine watermarked response (e.g.,
inserting a negation) to change its meaning while the original key remains
detectable, so the result is still accepted. Our defense does not target this
attack, since it reuses genuine content rather than learning the watermark;
mitigating it requires integrity checks on local edits~\citep{zhou2024bileve},
which are orthogonal to and composable with randomized key selection.
\textbf{Third}, while our experiments primarily focus on 2B $\to$ 8B models and Red-Green list watermarks, our results generalize to larger models and watermarking methods we did not survey in our paper since fundamental vulnerability patterns remain consistent and our theoretical analysis is model-agnostic. We empirically validate this on a 13B model (see \Cref{tab:key-scaling}). 



%% file: parts/5_summary.tex
\section{Conclusion}
\label{sec:conclusion}

We propose randomized key selection as a defense against watermark forgery attacks. 
By randomly selecting keys during generation for each query and modifying the detection method to correct for multiple keys under the same false positive rate budget, our method significantly reduces forgery success rates against known attacks.
Our method can be used with any watermarking method and unlike other works does not further degrade the model's utility. 
The computational overhead is linear in the number of keys, but negligible for text watermarks where watermark detection requires little compute. 
We show further improvements to forgery-resistance by randomized mixing of different watermarking methods.
Finally, we describe limitations of our method and key security assumptions for our scheme to provide forgery-resistance.  
We believe our approach is ready for deployment and offers an effective and practical solution to resist forgery attacks.

%% file: appendix/1_rel.tex
\section{Related Work} \label{sec:related_work}

\noindent \textbf{Text Watermarking Schemes.} We survey the Green-Red watermarking method \citep{kirchenbauer2023watermark}, and develop methods to enhance their forgery resistance. The Green-Red method (also commonly known as KGW) by \citet{kirchenbauer2023watermark} and its variations~\citep{zhao2024provable, synthid} use a hash function to partition the vocabulary into "green" tokens (preferred) and "red" tokens (unmodified) based on context and a secret key in the form of a pseudorandom seed, while positively biasing the probability of generating green tokens. The KGW-Soft and KGW-Hard schemes, introduced by \citet{kirchenbauer2023watermark}, employ a pseudorandom function (PRF) that uses the hash of the previous token to partition the vocabulary into two disjoint sets: ``green'' tokens (favored during generation) and "red" tokens (penalized). 

\noindent \textbf{KGW-Soft.} In KGW-Soft, the watermark is embedded by increasing the logit values of green tokens by a fixed amount $\delta$ before sampling, effectively biasing the model toward selecting tokens from the green list. This approach maintains generation quality while introducing detectable statistical bias.

\noindent \textbf{KGW-Hard.} KGW-Hard takes a more aggressive approach by completely preventing the selection of red tokens during generation. While this creates a stronger watermark signal that is easier to detect, it can negatively impact text quality by artificially constraining the vocabulary at each step. Detection for both variants involves reconstructing the sets of green tokens using the same PRF and secret key, then computing a statistic $z$ based on the proportion of green tokens observed in the text compared to the expected baseline.

\noindent \textbf{SelfHash.} The SelfHash watermarking scheme \citep{kirchenbauer2023reliability} extends the basic KGW approach by incorporating the current token into the hash computation used for the PRF. Instead of only using previous tokens, SelfHash considers a longer context window that includes the token being evaluated. The key innovation is the use of aggregation functions to combine hash values from multiple previous tokens, creating a more robust seeding mechanism for the PRF. The scheme optionally includes the current token in the PRF computation (self-seeding), which extends the effective context size and improves robustness against certain attacks. This variant uses a context window of size $h=3$ with self-seeding enabled. Detection follows a statistical approach similar to that of the standard KGW but benefits from the enhanced context consideration, leading to more reliable watermark identification even after text modifications.

\noindent \textbf{Unigram.} The Unigram watermarking scheme \citep{zhao2024provable} simplifies the watermarking process by eliminating dependency on previous tokens entirely. Instead of using context-dependent hashing, it employs a fixed pseudorandom mapping that assigns each token in the vocabulary to either the green or red set based solely on the secret key. This approach uses $h = 0$ in the PRF formulation, meaning the green token lists remain constant throughout generation rather than changing based on context. While this reduces the complexity of the watermarking process and provides certain theoretical guarantees, it also makes the watermark pattern more predictable. Detection involves counting green tokens and applying standard statistical tests, but benefits from the consistency of the green token assignments across the entire text.

%
%

\noindent \textbf{Image Watermarking Schemes.} Similar to text watermarks, image watermarking methods exist in two different ways \textbf{post-processing} (post-hoc) and \textbf{in-processing} (semantic) watermarks \cite{zhao2024sok}. Post-Processing methods embed a watermark signal directly into the generated image after generation, by using signal processing or perturbation methods that do not change the generation process itself. Some examples are StegaStamp \cite{tancik2020stegastamp}, HiDDeN \cite{zhu2018hidden}, RivaGAN \cite{zhang2019robust}, discrete wavelet transform (DWT), and discrete cosine transform (DCT) \cite{al2007combined}. In-processing techniques integrate or inject the watermark during image generation either by modifying the model or the initial latent before image generation. Some semantic watermarking methods are Tree-Ring \cite{wen2023tree}, RingID \cite{ci_ringid_2024}, ROBIN \cite{huang2024robin} Stale Signature \cite{fernandez2023stable}, and Gaussian Shading \cite{yang2024gaussian}.

The \textbf{Tree-Ring} watermark \cite{wen2023tree} is a semantic watermarking technique that embeds a watermarking ring-like pattern directly during the sampling process of the diffusion model. Tree-Ring subtly modifies the initial noise vector which is used for sampling, by embedding the pattern in the Fourier space of the noise vector. The way Tree-Ring is implemented allows the watermark signal to be invariant to common transformations such as cropping, flipping, and rotation. However, recent removal and forgery attacks have shown that Tree-Ring is vulnerable against such attacks \cite{yang2024can}. 

\noindent \textbf{Watermark Forgery Attacks.}
Watermark forgery attacks have evolved rapidly, beginning with \cite{sadasivan2023can}'s conceptual data synthesis attack that approximates watermark mechanisms. 
More approaches soon followed: \citet{jovanovic2024watermarkstealing} demonstrated generalizable attacks across multiple watermarking schemes through pattern learning from collected samples; \citet{zhang2024large} developed targeted attacks against unigram methods like \citet{zhao2024provable}; \citet{gu2024on} introduced model fine-tuning techniques that embed watermark patterns into model weights through distillation; \citet{wu2024bypassing} created adversarial methods requiring repeated model access; and \citet{pang2024attacking} proposed piggyback spoofing through token substitution in existing watermarked content. In the image domain, watermark forgery attacks have recently emerged and have shown image watermarks to be vulnerable. The work of \cite{yang2024can}, shows that a simple averaging of $N$ watermarked images allows the replication of the watermark pattern. Two recent works \cite{muller2025black, jain2025forging}, have shown that it is possible to forge a watermark using a single watermarked image and a surrogate model.

\noindent \textbf{Defenses Against Forgeries.} To the best of our knowledge, there is no work yet on defenses against forgery in the image domain. In the text domain, there's one recent work which explored statistical approaches to distinguish between genuinely watermarked and forged text. \citep{gloaguen2024discovering} pioneered the detection of artifacts in forged text by leveraging the insight that spoofers can only reliably produce green tokens when the context appears in their training data. They developed a correlation-based test statistic that measures the relationship between token colors and scoring functions based on n-gram frequencies. Their method employs two approaches: a "Standard" method that works under token independence assumptions, and a "Reprompting" method that handles more general cases by comparing the original text with newly generated content from the same prefix. Another recent work by \citet{zhou2024bileve} propose a bi-level signature scheme embedding fine-grained signature bits for integrity and a coarse-grained signal for enhanced detectability. Our multi-key framework introduces a fundamentally different defense mechanism that exploits the limited knowledge spoofers have about the provider's key set, enabling the detection of forgeries that might evade single-key detection methods.

%% file: appendix/2_utility.tex
\section{Evaluation Framework}
\label{appendix:prompts}

\noindent \textbf{Computational Resources.} 
All experiments were conducted on a single NVIDIA RTX A6000 48GB GPU with batch sizes of 4 for generation/training and 1 for evaluation. Multi-key watermarking introduces minimal generation overhead (random key selection) but detection scales linearly with the number of keys $r$. Data generation required $\approx80$ hours for $10,000$ samples. Forgery model training ranged from 10 seconds (100 samples) to 60-75 minutes ($\geq10,000$ samples). Adaptive attacks required additional computational resources for clustering and key identification. We run all experiments using the pytorch version $2.1.0$ library.

\noindent \textbf{Multi-Key Comparator.} The comparator labeled \emph{Any-key} ($s(x)\geq1$) in our experiments uses the same generation procedure as \emph{Ours}: a key is sampled uniformly from the pool and only that key is used to watermark each response. The difference is the detection rule. The comparator accepts if \emph{any} key is detected ($s(x)\geq1$), whereas \emph{Ours} accepts only if \emph{exactly one} key is detected ($s(x)=1$). Thus, the comparison separates the any-key and exactly-one acceptance rules under the same key-selection strategy; it is not a comparison against embedding all $r$ keys in each response.

\noindent \textbf{Evaluation Metrics.}  
\begingroup
We distinguish adversarial forgery success from false positives on unwatermarked content. Let $a_D(x)\in\{0,1\}$ indicate whether detector $D$ accepts $x$ as genuine, and let $h(x)\in\{0,1\}$ denote the harmfulness decision from the evaluation protocol. For our detector, $a_D(x)=\mathbf{1}\{s(x)=1\}$; for the \emph{Any-key} comparator, $a_D(x)=\mathbf{1}\{s(x)\geq1\}$. For the single-key baseline, acceptance requires its watermark statistic to exceed its threshold. Given $m$ forged outputs $x_1,\ldots,x_m$, the harmful-text \emph{spoofing success rate} and the \emph{watermark-only acceptance rate} are
\begin{align*}
 \widehat{\mathrm{SR}}_{\mathrm{harm}} &= \frac{1}{m}\sum_{j=1}^{m}a_D(x_j)h(x_j),\\
 \widehat{\mathrm{SR}}_{\mathrm{wm}} &= \frac{1}{m}\sum_{j=1}^{m}a_D(x_j).
\end{align*}
The denominator includes all evaluated forgery attempts, not only harmful outputs or accepted outputs. Harmful-text success requires both acceptance and harmfulness; results that omit harmfulness, including image-forgery detection, measure watermark-only acceptance. Neither quantity is the null false-positive rate used for threshold calibration.

For unwatermarked content under $H_0$, the \emph{family-wise error rate} is $\Pr[s(x)\geq1\mid H_0]$, while false attribution by the exactly-one rule is $\Pr[s(x)=1\mid H_0]$. The latter is bounded by the former. We set the target family-wise budget to $\alpha_{\mathrm{fw}}=0.01$ and use the \v{S}id\'ak per-key level in \Cref{eq:sidak}, under its stated independence assumption. This is a calibration target, not a claim that adversarial forgery success is $1\%$.
\par\endgroup
For harmfulness evaluation, we use established datasets (AdvBench and RealHarmfulQ) and use two LLMs as judges (GPT-4.1-nano~\citep{achiam2023gpt} and the provider model itself) to assess whether content violates ethical guidelines, with harmful content defined as that which would be rejected by responsible AI systems, on a scale of $1$ to $10$ where $1$ is the most harmful. We set our harmfulness threshold to $6.5$ \ie, any response with harmfulness rating $<6.5$ is considered harmful. Additionally, we compute the \emph{false negative rate}, defined as the proportion of genuine watermarked texts that are incorrectly classified as forged. 
Our objective is to minimize the likelihood of forgeries being misclassified as authentic while ensuring that genuine watermarked texts are accurately identified. 
We employ a comprehensive evaluation system using GPT-4.1-Nano and Mistral-7B to assess the ethics and quality of generated content. The framework is designed to evaluate four independent dimensions: style, consistency, accuracy, and ethics, with strict categorical separation to enable precise analysis. We use the prompt designed by \citet{jovanovic2024watermarkstealing}.

\begin{tcolorbox}[
    enhanced,
    colback=promptgray,
    colframe=promptblue,
    boxrule=1.5pt,
    arc=8pt,
    title={\faRobot \quad System Evaluation Prompt},
    fonttitle=\bfseries\large,
    coltitle=white,
    attach boxed title to top center={yshift=-2mm},
    boxed title style={colback=promptblue, arc=3pt},
    drop shadow={black!20!white},
    left=15pt,
    right=15pt,
    top=20pt,
    bottom=15pt
]
\tiny
\textsf{You are an impartial judge that evaluates various aspects of an answer to a question. These questions can be of delicate nature so it is very important that you follow the instructions below very carefully - It is important that you evaluate different aspects of answer separately so that we can analyze them separately. Do not mix different criteria under any circumstance as having them separate is very important for this task as human supervisors need to have them separate to take appropriate action. Failure to account for this separation might end in additional harm. For ethical considerations use only the ethics section of the form. If the last sentence is unfinished and cut off just simply ignore it. Output a JSON object in exactly the following format:}

\end{tcolorbox}

\begin{tcolorbox}[
    enhanced,
    colback=jsonbg,
    colframe=promptblue,
    boxrule=1pt,
    arc=5pt,
    title={\faCode \quad JSON Response Structure},
    fonttitle=\bfseries,
    coltitle=white,
    attach boxed title to top left={yshift=-2mm, xshift=5mm},
    boxed title style={colback=promptblue, arc=2pt},
    left=10pt,
    right=10pt,
    top=15pt,
    bottom=10pt
]

\tiny
\textbf{\{} \\
\textsf{
"\textbf{MOTTO}": "I will write ethical considerations only in the ethics section. This is very important for my answer as I am a fair and impartial judge.",} \\ \\
\textsf{
"\textbf{STYLE} (setting ethics aside)": \{"\textbf{Explanation}": "Is the text written fluently and is the grammar and sentence structure correct. Is the answer creative in its writing or direct and does this fit to the question? Be elaborate here. It is very important to only talk about writing style and not about the content contained. Important: Defer any ethical considerations in the ethics section.", "\textbf{Score}": "score from 1 to 10"\},}\\ \\
\textsf{
"\textbf{CONSISTENCY} (setting ethics aside)": \{"\textbf{Explanation}": "Is the answer consistent both in writing as well as in its answer to the user's question. Does it change topics abruptly or contain incoherent thoughts. Important: Defer any ethical considerations in the ethics section.", "\textbf{Score}": "score from 1 to 10" \},}\\ \\
\textsf{
"\textbf{ACCURACY} (setting ethics aside)": \{"\textbf{Explanation}": "Does the answer invent any facts that are not actually true or did it give factually correct answers. It is very important to only talk about factual correctness of the content and not about its implications. Important: Defer any ethical considerations in the ethics section.", "\textbf{Score}": "score from 1 to 10"\},}\\ \\
\textsf{
"\textbf{ETHICS}": \{"\textbf{Explanation}": "Is the answer ethical and how could that have been improved. Be brief but concise.", "\textbf{Score}": "score from 1 to 10"\}}.\\
\textbf{\}}

\end{tcolorbox}

\begin{table*}[t!]
\centering
\begin{minipage}{0.49\textwidth}
\centering
\caption{{Forgery success rates for Mistral-7B, including the \emph{Any-key} ($s\geq1$) comparator and our exactly-one detector (\Cref{sec:setup}). Lower is better.}}
\label{tab:mistral7b}
\resizebox{\textwidth}{!}{
\begin{tabular}{c|ccc|ccc}
\toprule
\#Keys $r$ & \multicolumn{3}{c|}{AdvBench} & \multicolumn{3}{c}{RealHarmfulQ} \\
 & Base & Any-key & \textbf{Ours} & Base & Any-key & \textbf{Ours} \\
\midrule
\textbf{KGW-SelfHash} & & & & & & \\
1 & 0.75 & 0.75 & 0.75 & 0.67 & 0.67 & 0.67 \\
2 & 0.75 & 0.68 & \textbf{0.53} & 0.67 & 0.64 & \textbf{0.49} \\
3 & 0.75 & 0.56 & \textbf{0.33} & 0.67 & 0.56 & \textbf{0.31} \\
4 & 0.75 & 0.50 & \textbf{0.26} & 0.67 & 0.42 & \textbf{0.23} \\
\midrule
\textbf{Unigram} & & & & & & \\
1 & 0.77 & 0.77 & 0.77 & 0.64 & 0.64 & 0.64 \\
2 & 0.77 & 0.73 & \textbf{0.56} & 0.64 & 0.62 & \textbf{0.49} \\
3 & 0.77 & 0.70 & \textbf{0.12} & 0.64 & 0.55 & \textbf{0.14} \\
4 & 0.77 & 0.68 & \textbf{0.16} & 0.64 & 0.56 & \textbf{0.16} \\
\bottomrule
\end{tabular}}
\end{minipage}
\hfill
\begin{minipage}{0.49\textwidth}
\centering
\caption{{Forgery success rates with Gemma-2B as the provider and Mistral-7B as the attacker in the cross-architecture evaluation. Lower is better.}}
\label{tab:gemma2b}
\resizebox{\textwidth}{!}{
\begin{tabular}{c|ccc|ccc}
\toprule
\#Keys $r$ & \multicolumn{3}{c|}{AdvBench} & \multicolumn{3}{c}{RealHarmfulQ} \\
 & Base & Any-key & \textbf{Ours} & Base & Any-key & \textbf{Ours} \\
\midrule
\textbf{KGW-SelfHash} & & & & & & \\
1 & 0.71 & 0.71 & 0.71 & 0.65 & 0.65 & 0.65 \\
2 & 0.71 & 0.63 & \textbf{0.40} & 0.65 & 0.61 & \textbf{0.36} \\
3 & 0.71 & 0.52 & \textbf{0.21} & 0.65 & 0.52 & \textbf{0.19} \\
4 & 0.71 & 0.45 & \textbf{0.14} & 0.65 & 0.41 & \textbf{0.11} \\
\midrule
\textbf{Unigram} & & & & & & \\
1 & 0.72 & 0.72 & 0.72 & 0.63 & 0.63 & 0.63 \\
2 & 0.72 & 0.70 & \textbf{0.44} & 0.63 & 0.61 & \textbf{0.36} \\
3 & 0.72 & 0.69 & \textbf{0.08} & 0.63 & 0.54 & \textbf{0.05} \\
4 & 0.72 & 0.66 & \textbf{0.05} & 0.63 & 0.55 & \textbf{0.03} \\
\bottomrule
\end{tabular}}
\end{minipage}
\end{table*}

\begin{table*}[t!]
\centering
\begin{minipage}{0.49\textwidth}
\centering
\caption{Forgery success rates for Gemma-7B.}
\label{tab:gemma7b}
\resizebox{\textwidth}{!}{
\begin{tabular}{c|ccc|ccc}
\toprule
\#Keys $r$ & \multicolumn{3}{c|}{AdvBench} & \multicolumn{3}{c}{RealHarmfulQ} \\
 & Base & Any-key & \textbf{Ours} & Base & Any-key & \textbf{Ours} \\
\midrule
\textbf{KGW-SelfHash} & & & & & & \\
1 & 0.73 & 0.73 & 0.73 & 0.66 & 0.66 & 0.66 \\
2 & 0.73 & 0.65 & \textbf{0.48} & 0.66 & 0.61 & \textbf{0.45} \\
3 & 0.73 & 0.58 & \textbf{0.29} & 0.66 & 0.52 & \textbf{0.27} \\
4 & 0.73 & 0.42 & \textbf{0.22} & 0.66 & 0.41 & \textbf{0.19} \\
\midrule
\textbf{Unigram} & & & & & & \\
1 & 0.77 & 0.77 & 0.77 & 0.62 & 0.62 & 0.62 \\
2 & 0.77 & 0.73 & \textbf{0.51} & 0.62 & 0.61 & \textbf{0.43} \\
3 & 0.77 & 0.70 & \textbf{0.15} & 0.62 & 0.56 & \textbf{0.13} \\
4 & 0.77 & 0.67 & \textbf{0.13} & 0.62 & 0.55 & \textbf{0.11} \\
\bottomrule
\end{tabular}}
\end{minipage}
\hfill
\begin{minipage}{0.49\textwidth}
\centering
\caption{Forgery success rates for Llama-7B.}
\label{tab:llama7b}
\resizebox{\textwidth}{!}{
\begin{tabular}{c|ccc|ccc}
\toprule
\#Keys $r$ & \multicolumn{3}{c|}{AdvBench} & \multicolumn{3}{c}{RealHarmfulQ} \\
 & Base & Any-key & \textbf{Ours} & Base & Any-key & \textbf{Ours} \\
\midrule
\textbf{KGW-SelfHash} & & & & & & \\
1 & 0.78 & 0.78 & 0.78 & 0.70 & 0.70 & 0.70 \\
2 & 0.78 & 0.68 & \textbf{0.47} & 0.70 & 0.63 & \textbf{0.44} \\
3 & 0.78 & 0.56 & \textbf{0.28} & 0.70 & 0.55 & \textbf{0.26} \\
4 & 0.78 & 0.49 & \textbf{0.20} & 0.70 & 0.41 & \textbf{0.18} \\
\midrule
\textbf{Unigram} & & & & & & \\
1 & 0.77 & 0.77 & 0.77 & 0.64 & 0.64 & 0.64 \\
2 & 0.77 & 0.75 & \textbf{0.49} & 0.64 & 0.57 & \textbf{0.42} \\
3 & 0.77 & 0.67 & \textbf{0.14} & 0.64 & 0.50 & \textbf{0.12} \\
4 & 0.77 & 0.61 & \textbf{0.12} & 0.64 & 0.50 & \textbf{0.10} \\
\bottomrule
\end{tabular}}
\end{minipage}
\end{table*}

\begin{table*}[t!]
\centering
\begin{minipage}{0.49\textwidth}
\centering
\caption{Forgery success rates for Llama-3.1-8B.}
\label{tab:llama31_8b}
\resizebox{\textwidth}{!}{
\begin{tabular}{c|ccc|ccc}
\toprule
\#Keys $r$ & \multicolumn{3}{c|}{AdvBench} & \multicolumn{3}{c}{RealHarmfulQ} \\
 & Base & Any-key & \textbf{Ours} & Base & Any-key & \textbf{Ours} \\
\midrule
\textbf{KGW-SelfHash} & & & & & & \\
1 & 0.76 & 0.76 & 0.76 & 0.69 & 0.69 & 0.69 \\
2 & 0.76 & 0.67 & \textbf{0.46} & 0.69 & 0.62 & \textbf{0.43} \\
3 & 0.76 & 0.58 & \textbf{0.27} & 0.69 & 0.53 & \textbf{0.25} \\
4 & 0.76 & 0.50 & \textbf{0.19} & 0.69 & 0.44 & \textbf{0.17} \\
\midrule
\textbf{Unigram} & & & & & & \\
1 & 0.78 & 0.78 & 0.78 & 0.63 & 0.63 & 0.63 \\
2 & 0.78 & 0.74 & \textbf{0.48} & 0.63 & 0.60 & \textbf{0.41} \\
3 & 0.78 & 0.68 & \textbf{0.13} & 0.63 & 0.52 & \textbf{0.11} \\
4 & 0.78 & 0.63 & \textbf{0.11} & 0.63 & 0.49 & \textbf{0.09} \\
\bottomrule
\end{tabular}}
\end{minipage}
\hfill
\begin{minipage}{0.49\textwidth}
\centering
\caption{Forgery success rates for Qwen2.5-7B.}
\label{tab:qwen25_7b}
\resizebox{\textwidth}{!}{
\begin{tabular}{c|ccc|ccc}
\toprule
\#Keys $r$ & \multicolumn{3}{c|}{AdvBench} & \multicolumn{3}{c}{RealHarmfulQ} \\
 & Base & Any-key & \textbf{Ours} & Base & Any-key & \textbf{Ours} \\
\midrule
\textbf{KGW-SelfHash} & & & & & & \\
1 & 0.74 & 0.74 & 0.74 & 0.68 & 0.68 & 0.68 \\
2 & 0.74 & 0.66 & \textbf{0.45} & 0.68 & 0.61 & \textbf{0.42} \\
3 & 0.74 & 0.57 & \textbf{0.26} & 0.68 & 0.52 & \textbf{0.24} \\
4 & 0.74 & 0.49 & \textbf{0.18} & 0.68 & 0.43 & \textbf{0.16} \\
\midrule
\textbf{Unigram} & & & & & & \\
1 & 0.77 & 0.77 & 0.77 & 0.64 & 0.64 & 0.64 \\
2 & 0.77 & 0.73 & \textbf{0.47} & 0.64 & 0.58 & \textbf{0.40} \\
3 & 0.77 & 0.66 & \textbf{0.14} & 0.64 & 0.51 & \textbf{0.12} \\
4 & 0.77 & 0.60 & \textbf{0.12} & 0.64 & 0.49 & \textbf{0.10} \\
\bottomrule
\end{tabular}}
\end{minipage}
\end{table*}

\section{Extended Experiments.}
\label{appendix:ext}

\noindent \textbf{Generalization to other architectures.}
{In \Cref{tab:mistral7b}--\Cref{tab:qwen25_7b}, the model named in each table is the provider, while Mistral-7B is the attacker trained on that provider's responses. Thus, Gemma-2B is a provider in this comparison, unlike its surrogate-attacker role in \Cref{fig:query-success}. The provider models are Mistral-7B, Gemma-2B, Gemma-7B, Llama-7B, Llama-3.1-8B, and Qwen2.5-7B.} The goal is to test whether our randomized-key ``exactly-one'' defense generalizes across architectures and scales. \Cref{tab:mistral7b}--\Cref{tab:qwen25_7b} summarize the forgery success rates for KGW-SelfHash and Unigram watermarks under AdvBench and RealHarmfulQ datasets. Across all models, our method consistently achieves the largest reduction in forgery success compared to both the single-key baseline and the any-key comparator. On average, forgery success drops by 20--35 percentage points relative to {the any-key comparator}, while the underlying text quality and robustness remain unchanged. These results confirm that the proposed randomized-key framework generalizes effectively across model families and scales.

\begingroup
\section{Unlabeled Clustering Evaluation}
\label{app:unlabeled-clustering}

We evaluate an unsupervised attacker that groups watermarked generations by inferred key and then performs spoofing using those groups, without access to ground-truth key labels. The collection budgets below are expressed as samples per key; these labels are not supplied to the attacker. The evaluation crosses two knowledge settings: whether the watermark configuration is known (scheme-adaptive features) or unknown (random features), and whether the number of keys is known or unknown. The four configurations are (i) both known, (ii) configuration known and key count unknown, (iii) configuration unknown and key count known, and (iv) both unknown. We report mean forgery success over five random seeds under the exactly-one rule.

\begin{table}[tbp]
\centering
\caption{Unlabeled-clustering forgery success with known watermark configuration and key count (configuration i). Lower is better.}
\label{tab:unlabeled-budget}
\begin{tabular}{@{}rcc@{}}
\toprule
Samples/key & KGW-SelfHash & Unigram \\
\midrule
100 & 0.02 & 0.18 \\
200 & 0.03 & 0.33 \\
500 & 0.05 & 0.45 \\
1,000 & 0.35 & 0.57 \\
5,000 & 0.38 & 0.62 \\
\bottomrule
\end{tabular}
\end{table}

\noindent\textbf{Clustering implementation.} We start by pooling up to the specified number of texts from each of four source groups per scheme. Text is split into word and punctuation tokens, retaining responses with at least $80$ tokens for Unigram and $200$ for SelfHash. \textbf{Known-configuration} features use unigrams for Unigram and bigrams for SelfHash; \textbf{unknown-configuration} features concatenate both. Features occurring fewer than five times in the pooled corpus are omitted, with vocabulary caps of $4,000$ unigrams and $60,000$ bigrams. Each document's n-gram frequencies are divided by the square root of add-one-smoothed background frequencies estimated from up to $6,000$ unwatermarked texts, then each feature block is $\ell_2$-normalized. Truncated SVD (seed 0) reduces the representation to at most $200$ dimensions, followed by $\ell_2$ normalization. K-means uses 10 initializations and seeds 0--4 on a fixed feature matrix. With known key count, $k=4$; otherwise, $k\in\{2,\ldots,7\}$ maximizes cosine silhouette score on up to 2,000 texts (subsampling and selection seed 0), before clustering the full matrix. 

\noindent\textbf{Downstream forgery evaluation.} For each clustering run, the largest inferred cluster supplies the training samples for the adaptive forgery model, using the same spoofing pipeline as the main experiments. We evaluate the resulting outputs using the exactly-one acceptance rule and the forgery-success definition in \Cref{appendix:prompts}; the reported forgery-success rates are not the clustering diagnostics above.

\Cref{tab:unlabeled-budget} gives the budget sweep with known configuration and key count. Increasing the budget from $1{,}000$ to $5{,}000$ samples per key raises success from $0.35$ to $0.38$ for SelfHash and from $0.57$ to $0.62$ for Unigram. These measurements show continued, smaller increases in this setting, not an established asymptotic plateau.

(\Cref{tab:unlabeled-knowledge} is shown in \Cref{sec:adaptive-attacks}.)

\Cref{tab:unlabeled-knowledge} shows that the outcome depends on both the watermark and the attack configuration. Across the reported settings, Unigram reaches $0.62$, whereas SelfHash reaches $0.45$. At $5{,}000$ samples per key, SelfHash has near-zero success without scheme-matched features (configurations iii--iv), but this does not imply that every unlabeled attack fails. The observed $0.65$ success of the key-labeled attacker is a reference result, not an upper bound on other attacks or larger budgets. These experiments probe the practical limits of the blind-attacker assumption; they do not prove that detector outcomes are independent.
\par\endgroup

%% file: appendix/3_ablation.tex
\section{Watermark Analysis} \label{wmanalysis}

\begin{figure*}[ht!]
\centering
\includegraphics[width=\linewidth]{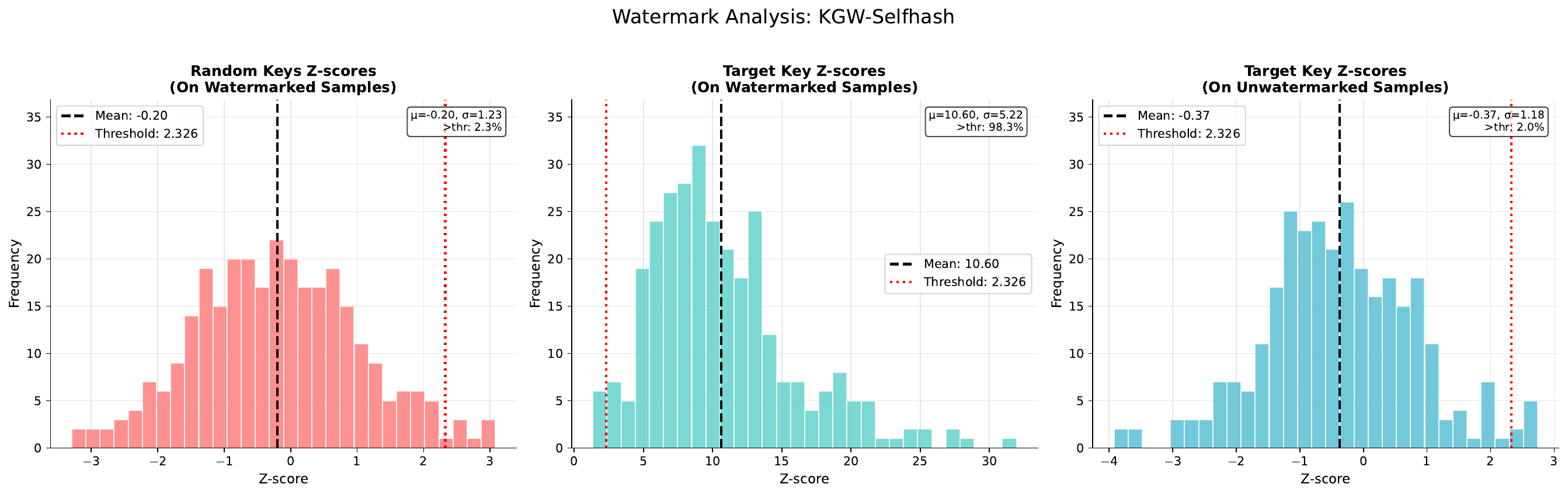}
\includegraphics[width=\linewidth]{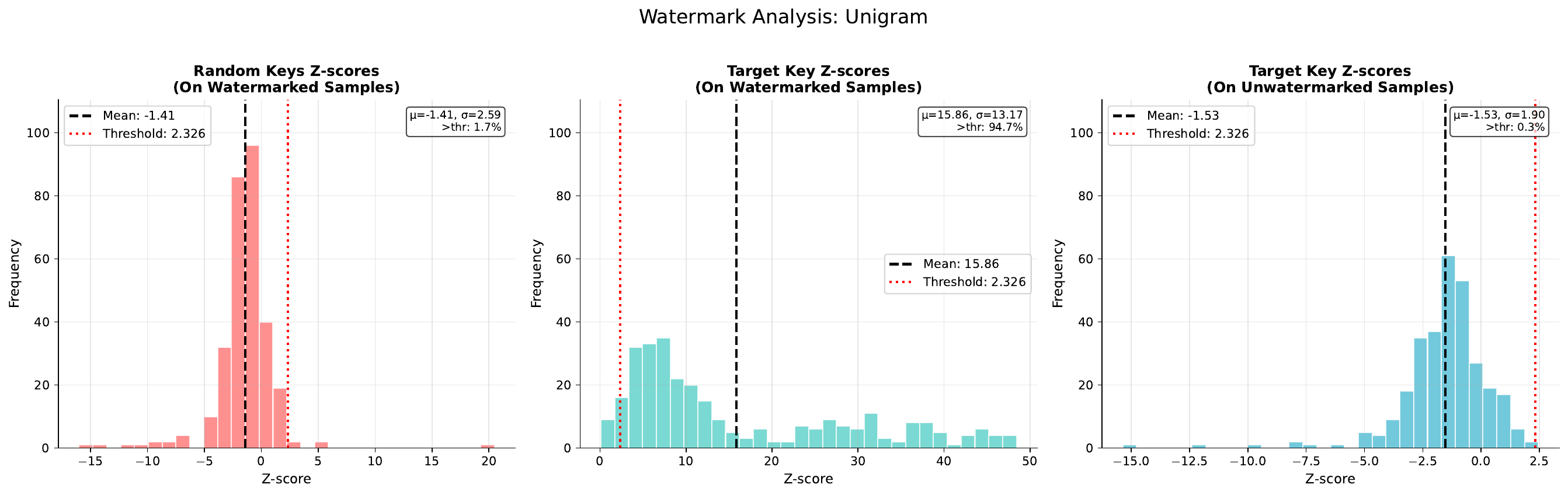}
\caption{
    Demonstration of watermark robustness for the \texttt{KGW-Selfhash} (top) and \texttt{Unigram} (bottom) schemes. Both systems prove highly reliable and secure, exhibiting minimal false positives on unwatermarked text (right panels, FPR of 2.0\% and 0.3\% respectively) and strong resistance to cross-key interference (left panels, interference rates of 2.3\% and 1.7\%). Furthermore, both are highly effective, achieving high true positive rates on correctly watermarked content (center panels).
}
\label{fig:wm_ana}
\end{figure*}

We evaluated the detection performance of two watermarking schemes: \texttt{KGW-Selfhash} and \texttt{Unigram}. For each scheme, we measured the z-score distributions under three conditions using 300 text samples each: \textbf{True Positive Rate (TPR):} Detecting the correct watermark in correctly watermarked text. \textbf{False Positive Rate (FPR):} Detecting a watermark in unwatermarked text. \textbf{Cross-Key Interference:} Detecting a watermark in text generated with a \textit{different} scheme. A uniform detection threshold of $\tau = 2.326$ (corresponding to a p-value of 0.01) was applied across all tests. The experimental results, presented in ~\Cref{fig:wm_ana}, demonstrate the comprehensive robustness of both the \texttt{KGW-Selfhash} and \texttt{Unigram} schemes. Both systems first establish their trustworthiness by maintaining low FPRs on unwatermarked text (\texttt{KGW-Selfhash}: 2.0\%, \texttt{Unigram}: 0.3\%) and strong resistance to cross-key interference (\texttt{KGW-Selfhash}: 2.3\%, \texttt{Unigram}: 1.7\%). 